\documentclass[a4paper,12pt]{article}
\usepackage{amsmath,enumerate}
\usepackage{graphicx}

\newcommand{\lapprox}{%
\mathrel{%
\setbox0=\hbox{$<$}
\raise0.6ex\copy0\kern-\wd0
\lower0.65ex\hbox{$\sim$}
}}
\newcommand{\gapprox}{%
\mathrel{%
\setbox0=\hbox{$>$}
\raise0.6ex\copy0\kern-\wd0
\lower0.65ex\hbox{$\sim$}
}}

\begin{document}

\begin{center}

{\Large \bf Implications of Higgs to diphoton decay rate in the bilinear
R-parity violating supersymmetric model}\\[20mm]

Raghavendra Srikanth Hundi\\
Centre for High Energy Physics,\\
Indian Institute of Science,\\
Bangalore 560 012, India.\\
E-mail: srikanth@cts.iisc.ernet.in \\[20mm]

\end{center}

\begin{abstract}
The Large Hadron Collider has recently discovered a Higgs-like particle
having a mass around 125 GeV and also indicated that there is an
enhancement in the Higgs to diphoton decay rate as compared to that
in the standard model. We have studied implications of these discoveries in
the bilinear R-parity violating supersymmetric model, whose main motivation
is to explain the non-zero masses for neutrinos. The R-parity violating
parameters in this model are $\epsilon$ and $b_\epsilon$, and these
parameters determine
the scale of neutrino masses. If the enhancement in the Higgs to diphoton
decay rate is true, then we have found $\epsilon\gapprox 0.01$ GeV and
$b_\epsilon\sim$ 1 GeV$^2$ in order to be compatible with the neutrino
oscillation data. Also, in the above mentioned analysis, we can determine
the soft masses of sleptons ($m_L$) and CP-odd Higgs boson mass ($m_A$).
We have estimated that $m_L\gapprox$ 300 GeV and $m_A\gapprox$ 700 GeV.
We have also commented on the allowed values of $\epsilon$ and
$b_\epsilon$, in case there is no
enhancement in the Higgs to diphoton decay rate.
Finally, we present a model to explain the smallness
of $\epsilon$ and $b_\epsilon$.
\end{abstract}


\newpage

\section{Introduction}

The ATLAS and CMS collaborations of Large Hadron Collider (LHC) have recently
discovered a bosonic particle whose mass being around 125 GeV \cite{LHC}.
The data from the LHC is strongly favouring the spin of this bosonic particle
to be zero and it is
consistent with the Higgs boson \cite{Higgs}, which is necessary to achieve
the electroweak symmetry breaking. The ATLAS and CMS
groups have analyzed the decay properties of this Higgs-like particle
into various standard model fields. An indication for the excess of events in
the Higgs to diphoton channel as compared to that in the standard model (SM)
has been reported. Explicitly, by defining the quantity
\begin{equation}
R_{\gamma\gamma}=\frac{\left[\sigma(pp\to h)\times {\rm BR}(h\to \gamma\gamma)
\right]_{\rm observed}}{\left[\sigma(pp\to h)\times {\rm BR}(h\to \gamma\gamma)
\right]_{\rm SM}},
\label{E:Rgg}
\end{equation}
where $h$ is the Higgs boson, the ATLAS and CMS had earlier reported that
$R_{\gamma\gamma}=$ $1.8\pm 0.5$ and $1.56\pm 0.43$ \cite{LHC}, respectively.
The above quoted values for $R_{\gamma\gamma}$ have been recently
updated in March 2013 at the conference
Rencontres de Moriond. The ATLAS group has claimed $R_{\gamma\gamma}
=1.65^{+0.34}_{-0.30}$ \cite{Hubaut}, which indicates a slight
enhancement in the $h\to\gamma\gamma$ channel. However, the CMS group has
reported that $R_{\gamma\gamma}$ could be $0.78^{+0.28}_{-0.26}$ or
$1.11^{+0.32}_{-0.30}$, depending on the type of the analysis \cite{Ochando}.
The values quoted by the CMS group imply that the discovered Higgs boson is
consistent with the SM within the uncertainties. We can hope that the
future analysis at ATLAS and CMS can resolve the differences in
$R_{\gamma\gamma}$. At this moment, it is worth to analyse by assuming
that the discovery made at the LHC favours new physics.

New physics has been motivated by several considerations and some of them
are gauge hierarchy problem and smallness of neutrino masses. Gauge
hierarchy problem can be solved by proposing supersymmetry \cite{susy}. In
supersymmetric models the Higgs boson mass can be around the electroweak
scale and it is protected from radiative corrections.
The weakly interacting neutrinos are found
to have non-zero masses which should not exceed 1 eV.
The non-zero masses for
neutrinos and upper limits on them have been established by
neutrino oscillation experiments \cite{osci}, cosmological observations
\cite{cosmo} and $\beta$-decay experiments \cite{beta}.
Since the neutrino masses should
be smaller than the electroweak scale by at least twelve orders of magnitude,
the smallness of their masses indicate a new mechanism for mass generation.

To solve both the gauge hierarchy problem and smallness of neutrino masses,
bilinear R-parity violating supersymmetric (BRPVS) model is a viable option.
For a review on the BRPVS model, see Ref. \cite{BRPVSmod}.
This model is a minimal extension of the minimal supersymmetric standard
model (MSSM). In the BRPVS model, additional bilinear terms of the forms
$\epsilon\hat{L}\hat{H}_u$ and $b_\epsilon \tilde{L}H_u$
are added to the superpotential and scalar potential
respectively.
Here $\hat{L}(\tilde{L})$ and $\hat{H}_u(H_u)$ are superfields
(scalar components)
of lepton and up-type Higgs doublets respectively.
The above mentioned bilinear terms violate lepton number and also
the R-parity.
$\epsilon$ is a mass parameter and $b_\epsilon$ is a mass-square parameter.
Provided that the $\epsilon$ and $b_\epsilon$ are very small, the masses
of neutrinos can be shown to be consistent with the observed neutrino
oscillation data
\cite{Hirsch-etal,Davidson-etal,GR}. One may
explain the smallness of $\epsilon$ and $b_\epsilon$ by proposing additional
symmetries \cite{Rpsmall} or by embedding this model in a high scale physics
\cite{HPT}.

The BRPVS model has rich phenomenology \cite{Rpheno,TestRp}.
In this work we want to study the affects of recent
discoveries at the LHC on the parameter space of the BRPVS model. As mentioned
above that the BRPVS model is an extension of MSSM and the additional
parameters in it are $\epsilon$ and $b_\epsilon$. Moreover, both $\epsilon$
and $b_\epsilon$ should be very small in order to account
for the smallness of neutrino masses. As a result of this,
in the BRPVS model, the contribution
to the Higgs boson mass and also to the quantity $R_{\gamma\gamma}$ are
dominantly determined by the MSSM parameters. In order to have light Higgs
boson mass $m_h\sim$ 125 GeV, the stop masses should be considerably high as
well as large mixing is needed in the stop sector \cite{Carena-etal,Cao-etal,
Arbey-etal}. However, parameters in the
squark sector do not affect the neutrino masses in the BRPVS model. On the
other hand, to have $R_{\gamma\gamma}>1$ it has been shown that relatively
light stau masses and large left-right mixing in the stau sector are required
\cite{Carena-etal}. Essentially,
this would mean that the soft parameters of slepton masses ($m_L$), higgsino
mass parameter ($\mu$) and the ratio of vacuum expectation values (vevs) of the
two neutral Higgs fields ($\tan\beta$) determine $R_{\gamma\gamma}$. We will
show that CP-odd Higgs boson mass ($m_A$) also has a role to play
in the enhancement of Higgs to diphoton decay rate. Shortly below we
will explain that the
parameters which determine $R_{\gamma\gamma}$ can affect
the neutrino masses in the BRPVS model. It is to remind that in the singlet
extension of MSSM, enhancement in $R_{\gamma\gamma}$
can be made not necessarily with light staus \cite{NMSSM}.

In the BRPVS model, one neutrino state acquires non-zero mass at tree
level due to mixing between flavor
neutrinos and neutralinos \cite{Davidson-etal,GR}. The remaining two
neutrino states acquire masses at 1-loop level due to mixing between
sneutrinos and the three neutral Higgs bosons \cite{Davidson-etal,GR}.
Explicitly, apart from $\epsilon$ and $b_\epsilon$, the neutrino masses
in this model are dominantly depended on
the neutralino parameters ($M_{1,2}$, $\mu$, $\tan\beta$), $m_L$ and $m_A$.
From the discussion in the previous paragraph,
we can understand that the parameters which determine the neutrino masses
in the BRPVS model have a role to play in the enhancement of Higgs to
diphoton decay rate. From this perspective, we can understand
that the requirement
of $R_{\gamma\gamma}>1$ can lead to certain allowed values for $\epsilon$
and $b_\epsilon$, which determine the overall scales of neutrino masses.

Both the ATLAS and CMS groups of the LHC are yet to confirm whether
$R_{\gamma\gamma}>1$ or not. Hence we have also analyzed the case
$R_{\gamma\gamma}\leq 1$. In either of these cases we will see that
the allowed values of $\epsilon$ and $b_\epsilon$
are small, and their smallness can be motivated from a high scale physics.
While motivating these parameters from a high scale physics, we can
also predict allowed ranges for $m_L$, $m_A$ and also about other
supersymmetric parameters.

The paper is organized as follows. In the next section, we give a brief
overview of the BRPVS model and also describe the neutrino masses in this
model. In the same section we will also explain the
relevant quantities regarding the Higgs boson mass and $R_{\gamma\gamma}$.
In Sec. 3, we describe our results on $\epsilon$ and $b_\epsilon$ which
are compatible with neutrino oscillation data and also with
$R_{\gamma\gamma}$. We then motivate these results from a high
scale physics. We conclude in Sec. 4.

\section{The BRPVS model}

The superpotential of the BRPVS model is
\begin{equation}
W=Y_u^{ij}\hat{Q}_i\hat{U}_j\hat{H}_u - Y_d^{ij}\hat{Q}_i\hat{D}_j\hat{H}_d
- Y_e^{ij}\hat{L}_i\hat{E}_j\hat{H}_d + \mu\hat{H}_u\hat{H}_d
+ \epsilon_i\hat{L}_i\hat{H}_u,
\label{E:supbi}
\end{equation}
where the indices $i,j$ run from 1 to 3. The superfields
$\hat{Q}$, $\hat{U}$ and $\hat{D}$ are doublet,
singlet up-type and singlet down-type quark fields,
respectively. $\hat{L}$ and $\hat{E}$ are doublet
and singlet charged lepton superfields, respectively. $\hat{H}_u$
and $\hat{H}_d$ are up- and down-type Higgs superfields,
respectively.
As already explained in the previous section,
the superpotential terms $\hat{L}_i\hat{H}_u$ are the bilinear R-parity
violating terms. The addition of these R-parity violating terms makes the
superpotential of BRPVS model to differ from that of MSSM. Another difference
between the BRPVS model and the MSSM is that the soft scalar potential
in the BRPVS model has additional terms which correspond to the
$\hat{L}_i\hat{H}_u$-terms. The form of the soft scalar potential in the
BRPVS model is
\begin{eqnarray}
V_{\rm soft}^{\rm BRPVS}&=& V_{\rm soft}^{\rm MSSM} + 
\left[(b_\epsilon)_i\tilde{L}_iH_u + {\rm c.c.}\right],
\label{E:vrp}
\\
V_{\rm soft}^{\rm MSSM} &=& \frac{1}{2}\left(M_1\tilde{B}\tilde{B}+M_2
\tilde{W}\tilde{W}+M_3\tilde{g}\tilde{g}+{\rm c.c}\right)+
m_{H_u}^2H^*_uH_u+m_{H_d}^2H^*_dH_d+
\nonumber \\
&&+(m_Q^2)_{ij}\tilde{Q}^*_i\tilde{Q}_j + (m_U^2)_{ij}\tilde{U}^*_i\tilde{U}_j
+(m_D^2)_{ij}\tilde{D}^*_i\tilde{D}_j+
(m_L^2)_{ij}\tilde{L}^*_i\tilde{L}_j + (m_E^2)_{ij}\tilde{E}^*_i\tilde{E}_j
\nonumber \\
&&+\left[(A_U)_{ij}\tilde{Q}_i\tilde{U}_jH_u+ (A_D)_{ij}\tilde{Q}_i\tilde{D}_jH_d+ (A_E)_{ij}\tilde{L}_i\tilde{E}_jH_d
+b_\mu H_uH_d + {\rm c.c.}\right].
\label{E:vmssm}
\end{eqnarray}
The explicit form of the soft terms in the
MSSM are given in the form of $V_{\rm soft}^{\rm MSSM}$.

\subsection{Neutrino masses in the BRPVS model}

In this subsection we will describe the neutrino masses, which are
generated mainly due to the bilinear R-parity violating terms. In fact,
these bilinear terms violates lepton number, and as a result,
the sneutrinos can acquire non-zero vevs. However, without loss of generality,
we work in a particular basis where the vevs of sneutrinos are kept to
be zero.

The $\epsilon$-term of Eq. (\ref{E:supbi}) generate mixing between
flavor neutrinos ($\nu_i$) and higgsino. In a basis where
$\psi_N = (\tilde{B},
\tilde{W}^3,\tilde{H}_u^0,\tilde{H}_d^0,\nu_1,\nu_2,\nu_3)^T$, at the
tree level we get the following mixing masses:
${\cal L} = -\frac{1}{2}\psi^T_N M_N \psi_N + {\rm h.c.}$,
where
\begin{equation}
M_N=
\left(\begin{array}{cc}
M_{\chi^0} & m \\
m^T & 0
\end{array}\right),
\label{eq:n0}
\end{equation}
\begin{equation}
M_{\chi^0}=\left(\begin{array}{cccc}
M_1 & 0 & \frac{1}{\sqrt{2}}g_1v_u & -\frac{1}{\sqrt{2}}g_1v_d \\
0 & M_2 & -\frac{1}{\sqrt{2}}g_2v_u & \frac{1}{\sqrt{2}}g_2v_d \\
\frac{1}{\sqrt{2}}g_1v_u & -\frac{1}{\sqrt{2}}g_2v_u & 0 & -\mu \\
-\frac{1}{\sqrt{2}}g_1v_d & \frac{1}{\sqrt{2}}g_2v_d & -\mu & 0
\end{array}\right),
\quad
m = \left(\begin{array}{ccc}
0 & 0 & 0 \\
0 & 0 & 0 \\
\epsilon_1 & \epsilon_2 & \epsilon_3 \\
0 & 0 & 0
\end{array}\right).
\label{eq:mchim}
\end{equation}
Here, $g_1,g_2$ are the gauge couplings corresponding to the gauge groups
U(1)$_Y$ and SU(2)$_L$, respectively.
The vevs of Higgs scalar fields are defined as:
$\langle H_d^0 \rangle = v_d = v\cos\beta,\langle H_u^0 \rangle = v_u
= v\sin\beta$,
where $v=174$ GeV is the electroweak scale. Assuming that $\epsilon_i$
are very small compared to the TeV scale, at leading order, after integrating
away the components of neutralinos, we get the neutrino mass matrix as
$m_\nu = -m^T M^{-1}_{\chi^0} m$. However, this leading neutrino mass matrix
will give only one non-zero mass eigenvalue \cite{Davidson-etal,GR}.
In a realistic scenario we need at least two non-zero mass eigenvalues
for neutrinos \cite{osci}. It can be shown that the other two neutrino
states get non-zero masses due to radiative contributions
\cite{Davidson-etal,GR}. At 1-loop level, neutrinos get non-zero masses
because of mixing between sneutrinos and neutral Higgs boson states
\cite{Davidson-etal,GR}, and this mixing is driven by the
soft $b_\epsilon$-term of Eq. (\ref{E:vrp}).
It has been shown in Ref. \cite{Davidson-etal,GR} that at 1-loop level,
the dominant
contribution to neutrino masses come from diagrams involving two insertions
of $b_\epsilon$, provided the tree level mass eigenvalue is dominant over the
loop contribution. Based on this, below we present the complete expression
for neutrino mass matrix. In this expression we assume degenerate
masses for sneutrinos.

The neutrino mass matrix in the BRPVS model, up to leading contributions,
can be written as \cite{Davidson-etal,GR}
\begin{equation}
(m_\nu)_{ij}=a_0\epsilon_i\epsilon_j+a_1(b_\epsilon)_i(b_\epsilon)_j,
\label{E:tl1l}
\end{equation}
where the indices $i,j$ run from 1 to 3. The first term in the
above equation is due to the tree level effect, which is described
in the previous paragraph, and the second term
is from 1-loop diagrams. The expressions for $a_0$ and $a_1$ are
\cite{Davidson-etal,GR}
\begin{eqnarray}
a_0&=&\frac{m_Z^2m_{\tilde{\gamma}}\cos^2\beta}{\mu(m_Z^2m_{\tilde{\gamma}}
\sin 2\beta-M_1M_2\mu)},\quad m_{\tilde{\gamma}}=\cos^2\theta_W M_1+\sin^2\theta_W M_2,
\nonumber \\
a_1&=&\sum_{i = 1}^{4}\frac{(g_2(U_0)_{2i}-g_1(U_0)_{1i})^2}
{4\cos^2\beta}(m_{N^0})_i\left(I_4(m_h,m_{\tilde \nu},m_{\tilde \nu},(m_{N^0})_i)\cos^2(\alpha-\beta)
\right.
\nonumber \\
&& \left. +I_4(m_H,m_{\tilde \nu},m_{\tilde \nu},(m_{N^0})_i)\sin^2(\alpha-\beta)
-I_4(m_A,m_{\tilde \nu},m_{\tilde \nu},(m_{N^0})_i)\right),
\label{E:a0a1}
\end{eqnarray}
where $m_Z$ is the $Z$ boson mass, $\theta_W$ is the Weinberg angle,
and the $m_h$, $m_H$ and $m_A$ are the light, heavy and
pseudo-scalar Higgs boson masses, respectively. The unitary matrix $U_0$
diagonalizes the neutralino mass matrix as $(U_0^{\rm T}M_{\chi^0}U_0)_{ij}
=(m_{N^0})_i\delta_{ij}$, where $(m_{N^0})_i$ are the neutralino
mass eigenvalues. $m_{\tilde \nu}$ is the mass of sneutrino
field. $\alpha$ is the mixing angle
in the CP-even Higgs sector. The function $I_4$ is given by
\begin{eqnarray}
I_4(m_1,m_2,m_3,m_4)&=&\frac{1}{m_1^2-m_2^2}[I_3(m_1,m_3,m_4)-
I_3(m_2,m_3,m_4)],
\nonumber \\
I_3(m_1,m_2,m_3)&=&\frac{1}{m_1^2-m_2^2}[I_2(m_1,m_3)-I_2(m_2,m_3)],
\nonumber \\
I_2(m_1,m_2)&=&-\frac{1}{16\pi^2}\frac{m_1^2}{m_1^2-m_2^2}
\ln\frac{m_1^2}{m_2^2}.
\label{eq:i4}
\end{eqnarray}

Since we have assumed degenerate masses for sneutrinos, the neutrino matrix in
Eq. (\ref{E:tl1l}) generate only two non-zero masses, which is sufficient
to explain the solar and atmospheric neutrino mass scales \cite{osci}.
By taking the supersymmetric mass parameters to be few 100 GeV in
$a_0$ and $a_1$ of Eq. (\ref{E:tl1l}), we can estimate the magnitudes
of the unknown parameters
$\epsilon_i$ and $(b_\epsilon)_i$, in order to have a neutrino mass scale
of 0.1 eV. Taking into account of the partial cancellations of Higgs boson
contributions in $a_1$ \cite{GR}
of Eq. (\ref{E:tl1l}), for $\tan\beta\sim 10$, we get
$\epsilon_i\lapprox 10^{-3}$ GeV and
$(b_\epsilon)_i\sim$ 1 GeV$^2$.
As already described before, the estimated magnitudes of $\epsilon$ and
$b_\epsilon$ are very small in order to explain the smallness of
neutrinos masses,
and in this work, we analyze if these estimated magnitudes are compatible
with the Higgs to diphoton decay rate measured at the LHC.

\subsection{Higgs to diphoton decay in the BRPVS model}

We have already explained before that the BRPVS model is an extension
of MSSM, where the additional terms are
$\epsilon$- and $b_\epsilon$-terms of
Eqs. (\ref{E:supbi}) and (\ref{E:vrp}), respectively. We have argued before
that both the parameters $\epsilon$ and $b_\epsilon$ need to be very small
in order to explain the smallness of neutrino masses.
As a result of this, in the BRPVS model, the masses
and decay widths of Higgs boson states are almost same as that in the MSSM.
The leading contribution to light Higgs boson mass up to 1-loop level is
given by \cite{Djouadi}
\begin{equation}
m_h^2=m_Z^2\cos^22\beta+\frac{3m_t^4}{4\pi^2v^2}\ln\frac{M_S^2}{m_t^2}
+\frac{3m_t^4X_t^2}{4\pi^2v^2M_S^2}\left(1-\frac{X_t^2}{12M_S^2}\right),
\end{equation}
where $m_t$ is the top quark mass, $M_S=\sqrt{m_{\tilde{t}_1}m_{\tilde{t}_2}}$
and $X_t=A_t-\mu /\tan\beta$, $A_t=(A_U)_{33}$. Here, $m_{\tilde{t}_{1,2}}$
are masses of the stops. The second and third terms in the above equation
arise due to 1-loop corrections from top and stops. The tree level contribution
to $m_h$ is $\approx$ 91 GeV and in order to have $m_h\sim$ 125 GeV, the
loop contributions from top and stops should be substantially large.
As a result
of this, the light Higgs boson mass is dominantly determined by
parameters in the squark sector and the top mass. These parameters do not play
any role in determining the neutrino masses in the BRPVS model. However,
to be consistent with the recent Higgs boson mass of $\sim$ 125 GeV,
the above mentioned parameters should be fixed accordingly in the BRPVS model.
It is to remind that the loop contribution from top and stop would be maximum
if $X_t=\sqrt{6}M_S$. This choice of parameter space is known as
maximal mixing scenario \cite{Djouadi}. In our analysis, which will be
discussed below, we have considered
the maximal mixing scenario in order to have $m_h\sim$ 125 GeV.

On the other hand, in order to have enhancement in the Higgs to diphoton
decay rate, the
quantity defined in Eq. (\ref{E:Rgg}) has a role to play on neutrino
masses. Since the dominant production for light Higgs boson at the LHC takes
place through gluon fusion process, we reformulate $R_{\gamma\gamma}$ as
\begin{equation}
R_{\gamma\gamma}\approx
\frac{\left[\Gamma(h\to gg)\times {\rm BR}(h\to \gamma\gamma)
\right]_{\rm MSSM}}{\left[\Gamma(h\to gg)\times {\rm BR}(h\to \gamma\gamma)
\right]_{\rm SM}}.
\label{E:Rmssm}
\end{equation}
In the above expression, we have used $\sigma(gg\to h)$ to be proportional
to the decay width $\Gamma(h\to gg)$.
In the MSSM, the supersymmetric contribution to $\Gamma(h\to gg)$ is
from squarks, while
the decay width of $h\to\gamma\gamma$ gets supersymmetric contribution from
squarks, charged sleptons, charged Higgs bosons and charginos. For complete
expressions, up to leading order, for $\Gamma(h\to gg)$ and
$\Gamma(h\to\gamma\gamma)$ in the SM as well as in MSSM, see
Ref. \cite{Djouadi}.

A scan of parameter space in the MSSM has been
done in \cite{Carena-etal} and it has been reported that
to have $R_{\gamma\gamma}>1$ the masses of staus should be light and
the left-right mixing in the stau sector should be large. We will show later
that $R_{\gamma\gamma}$ has some sensitivity to the CP-odd Higgs boson mass.
The masses and mixing in the stau sector are determined
by the parameters $m_L^2$, $m_R^2$, $A_E$, $\mu$ and $\tan\beta$.
From the previous subsection, we can notice that the magnitudes of
$\epsilon$ and $b_\epsilon$ fix the neutrino mass eigenvalues in the
BRPVS model. Apart from this, the tree level neutrino masses are depended
on the neutralino parameters. Also, the 1-loop contribution
to neutrino masses are determined by the masses of neutralinos, sneutrinos
and neutral Higgs bosons. It is to be noticed that the sneutrino masses
are determined by the soft parameter $m_L^2$.

In the previous paragraph we have explained how the neutrino
masses in the BRPVS model are correlated with the $R_{\gamma\gamma}$.
We have studied this correlation and in the next section we present our
results. Apart from this correlation, one may also study additional bounds
arising from vacuum stabilization \cite{Kita}, which we leave it
for future studies.

\section{Results}

In this section we present our results on the correlation between
neutrino masses and $R_{\gamma\gamma}$ in the BRPVS model. We divide
this section into three parts. In Sec. 3.1 we describe the diagonalization
procedure of the neutrino mass matrix of Eq. (\ref{E:tl1l}), from which
we obtain expressions for neutrino mass eigenvalues in terms of model
parameters. In Sec. 3.2 we illustrate our method of calculating
the $R_{\gamma\gamma}$ by varying the model parameters. After scanning
over model parameters, we can obtain
the allowed parameter space of the BRPVS model, in order for the
neutrino oscillation data to be compatible with $R_{\gamma\gamma}>1$.
As stated before that the LHC has not yet confirmed $R_{\gamma\gamma}>1$,
so we make brief comments about the possibility of $R_{\gamma\gamma}\leq 1$.
From our numerical results we can see that the allowed values for
$\epsilon$ and $b_\epsilon$ should be very small.
We try to motivate the smallness of these
values from a high scale physics, which we describe in Sec. 3.3.

\subsection{Neutrino mass eigenvalues}

After diagonalizing the neutrino mass matrix of Eq. (\ref{E:tl1l}), we
should obtain mass eigenvalues as well as the mixing angles. The mixing
angles are incorporated in the well known Pontecorvo-Maki-Nakagawa-Sakata
unitary matrix, $U_{\rm PMNS}$, which is the diagonalizing matrix of
Eq. (\ref{E:tl1l}). We parametrize the $U_{\rm PMNS}$ as it is suggested
in \cite{pdg}. Among the three neutrino mixing angles,
$\theta_{12}$ and $\theta_{23}$
are found to be large, whereas, the third mixing angle $\theta_{13}$
is non-zero and is relatively small \cite{theta13}. From the recent
global fit to various neutrino oscillation data \cite{glob-fit}, we
can still choose
tri-bimaximal values for $\theta_{12}$ and $\theta_{23}$ \cite{tribi}.
Hence, we take
$\sin\theta_{12}=\frac{1}{\sqrt{3}}$, $\sin\theta_{23}=\frac{1}{\sqrt{2}}$.
As for the $\sin\theta_{13}$, at $3\sigma$ level, its fitted value
can be between
0.13 to 0.18 \cite{glob-fit}. In our analysis we take $\sin\theta_{13}$
to be anywhere in this $3\sigma$ range. For simplicity, we assume
the Dirac CP-odd phase, $\delta$, and the Majorana phases to be zero.

From the diagonalization of mass matrix in Eq. (\ref{E:tl1l}), we obtain
the following relation
\begin{equation}
m_\nu = U^*_{\rm PMNS}m_\nu^{\rm D}U^\dagger_{\rm PMNS},
\label{E:matrel}
\end{equation}
where $m_\nu^{\rm D}= {\rm Diag}(m_1,m_2,m_3)$ and
$m_{1,2,3}$ are the mass eigenvalues of neutrinos. For a given set
of neutrino mass eigenvalues and mixing angles, the above matrix equation
can be solved, since it involves
6 relations in terms of 6 unknown parameters
($\epsilon_i$, $(b_\epsilon)_i$). One possible solution to the above
matrix equation is given in \cite{Hundi}, in the limit of $s_{13}\equiv
\sin\theta_{13}=0$. Since,
now it has been established that $s_{13}\neq 0$ \cite{theta13}, below
we describe an approximate way of solving the above matrix relation.
Although $s_{13}\neq 0$, from the previous paragraph we can see that
$s_{13}\sim$ 0.1 and hence higher powers of $s_{13}$ are at least
one order of magnitude smaller than that of $s_{13}$.
Based on this observation, we can expand $\cos\theta_{13}=\sqrt{1-s_{13}^2}
\approx 1-\frac{1}{2}s_{13}^2+\cdots$. Using this expansion and also
fixing the $\theta_{12}$ and $\theta_{23}$ to their tri-bimaximal values,
we can expand $U_{\rm PMNS}$ in the following way.
\begin{eqnarray}
U_{\rm PMNS}&=& U_0+U_1s_{13}+U_2s_{13}^2+\cdots,
\nonumber \\
U_0&=&\left(\begin{array}{ccc}
\sqrt{\frac{2}{3}} & \frac{1}{\sqrt{3}} & 0 \\
-\frac{1}{\sqrt{6}} & \frac{1}{\sqrt{3}} & \frac{1}{\sqrt{2}} \\
\frac{1}{\sqrt{6}} & -\frac{1}{\sqrt{3}} & \frac{1}{\sqrt{2}}
\end{array}\right),\quad
U_1=\left(\begin{array}{ccc}
0 & 0 & 1 \\
-\frac{1}{\sqrt{3}} & -\frac{1}{\sqrt{6}} & 0 \\
-\frac{1}{\sqrt{3}} & -\frac{1}{\sqrt{6}} & 0
\end{array}\right),
\nonumber \\
U_2&=&\left(\begin{array}{ccc}
-\frac{1}{\sqrt{6}} & -\frac{1}{\sqrt{12}} & 0 \\
0 & 0 & -\frac{1}{\sqrt{8}} \\
0 & 0 & -\frac{1}{\sqrt{8}}
\end{array}\right).
\end{eqnarray}
From the above expansion of $U_{\rm PMNS}$, we can realize that
the right hand side of Eq. (\ref{E:matrel}) can be expressed as a power
series in terms of $s_{13}$. Such a matrix relation in Eq. (\ref{E:matrel})
can be solved if we also express the left hand side of it in a similar
power series expansion. Hence, we may propose the following series
expansions for $\epsilon_i$ and $(b_\epsilon)_i$.
\begin{eqnarray}
\epsilon_i&=&\epsilon_{i,0}+\epsilon_{i,1}s_{13}+\epsilon_{i,2}s_{13}^2+
\cdots,
\nonumber \\
(b_\epsilon)_i&=&(b_\epsilon)_{i,0}+(b_\epsilon)_{i,1}s_{13}+
(b_\epsilon)_{i,2}s_{13}^2+\cdots.
\label{E:powexp}
\end{eqnarray}
Here, we can assume that the coefficients $\epsilon_{i,0}$, $\epsilon_{i,1}$,
$\epsilon_{i,2}$,
etc in the expansion for $\epsilon_i$ have same order of magnitude to
one another. This also applies to
the coefficients in the power series expansion for $(b_\epsilon)_i$. Below
we show one solution for Eq. (\ref{E:matrel}), where we solve $\epsilon_i$
and $(b_\epsilon)_i$ up to ${\cal O}(s_{13})$.

Plugging Eq. (\ref{E:powexp}) in Eq. (\ref{E:matrel}), up to ${\cal O}(s_{13})$,
we get the following relations.
\begin{eqnarray}
\left[U_0m_\nu^{\rm D}U_0^{\rm T}\right]_{ij}
&=&a_0\epsilon_{i,0}\epsilon_{j,0}+a_1(b_\epsilon)_{i,0}(b_\epsilon)_{j,0}),
\label{E:leadrel}
\\
\left[U_0m_\nu^{\rm D}U_1^{\rm T}+
U_1m_\nu^{\rm D}U_0^{\rm T}\right]_{ij}&=& a_0(\epsilon_{i,0}
\epsilon_{j,1}
+\epsilon_{i,1}\epsilon_{j,0})+a_1((b_\epsilon)_{i,0}(b_\epsilon)_{j,1}
+(b_\epsilon)_{i,1}(b_\epsilon)_{j,0}).
\label{E:subrel}
\end{eqnarray}
One solution to the matrix relation in Eq. (\ref{E:leadrel}) is given
below \cite{Hundi}
\begin{eqnarray}
&& \epsilon_{1,0}=0,\epsilon_{2,0}=\epsilon_{3,0}=\epsilon,\quad
(b_\epsilon)_{1,0}=(b_\epsilon)_{2,0}=-(b_\epsilon)_{3,0}
=b_\epsilon,
\nonumber \\
&& m_1=0,\quad m_2=3a_1(b_\epsilon)^2,\quad m_3=2a_0\epsilon^2.
\label{E:leadsol}
\end{eqnarray}
As already described before, we can understand that $m_3$ and $m_2$
are determined by tree level and 1-loop level contributions, respectively,
to the neutrino masses
in the BRPVS model. The mass eigenvalue $m_1$ has come out be zero, since
we have assumed degenerate masses for sneutrinos. Using the solution at
leading order in Eq. (\ref{E:leadsol}), we can reduce the six independent
relations of Eq. (\ref{E:subrel}) into five, which are shown below.
\begin{eqnarray}
&& (b_\epsilon)_{1,1}=0,\quad (b_\epsilon)_{2,1}=(b_\epsilon)_{3,1},\quad
\epsilon_{2,1}=-\epsilon_{3,1},
\nonumber \\
&&-\frac{m_2-3m_3}{3\sqrt{2}}=a_0\epsilon\epsilon_{1,1}+
a_1b_\epsilon(b_\epsilon)_{2,1},\quad
-\frac{m_2}{3\sqrt{2}}=a_0\epsilon\epsilon_{2,1}+
a_1b_\epsilon(b_\epsilon)_{2,1}
\label{E:5rel}
\end{eqnarray}
The last two relations of Eq. (\ref{E:5rel}) can be solved for infinitesimally
many possible values of $\epsilon_{1,1}$, $\epsilon_{2,1}$ and
$(b_\epsilon)_{2,1}$. We obtain one simple solution by choosing
$\epsilon_{2,1} = 0$. As a result of this, for the given set of neutrino
mixing angles which we have described above,
a solution to the matrix relation of Eq. (\ref{E:matrel}),
solved up to ${\cal O}(s_{13})$, is
\begin{eqnarray}
&&\epsilon_1=\epsilon [\sqrt{2}s_{13}+{\cal O}(s_{13}^2)],\quad
\epsilon_2=\epsilon [1+{\cal O}(s_{13}^2)],\quad
\epsilon_3=\epsilon [1+{\cal O}(s_{13}^2)],\quad
(b_\epsilon)_1=b_\epsilon [1+{\cal O}(s_{13}^2)],
\nonumber \\
&&(b_\epsilon)_2=b_\epsilon [1-\frac{1}{\sqrt{2}}s_{13}+{\cal O}(s_{13}^2)],\quad
(b_\epsilon)_3=b_\epsilon [-1-\frac{1}{\sqrt{2}}s_{13}+{\cal O}(s_{13}^2)].
\end{eqnarray}
Here, $\epsilon$ and $b_\epsilon$ determine the non-zero neutrino
mass eigenvalues of the BRPVS model, which are given in Eq. (\ref{E:leadsol}).
We believe the procedure described above can be extended to solve
$\epsilon_i$ and $(b_\epsilon)_i$ up to second and higher order powers of
$s_{13}$.

\subsection{Computation of Higgs to diphoton decay rate}

As already explained that the Higgs to diphoton decay rate and the masses of
scalar Higgs bosons in the BRPVS model are almost
same as that in the MSSM. The enhancement related to this decay rate, as
quantified in Eq. (\ref{E:Rmssm}), and also the masses of Higgs bosons
have been computed with the HDECAY code \cite{Hdecay}. In our
numerical analysis, we have fixed off-diagonal elements of soft mass-squared
and $A$-terms of Eq. (\ref{E:vmssm}) to be zero, which is also incorporated
in the HDECAY code. In order to have the light Higgs
boson mass to be around 125 GeV, we have fixed
$(m_Q^2)_{33}=(m_U^2)_{33}=(m_D^2)_{33}= (800~{\rm GeV})^2$,
$(A_U)_{33}=\sqrt{6(m_Q^2)_{33}}+\mu\cot\beta$ and $(A_D)_{33}=0$. The specific
choice for $(A_U)_{33}$ has been motivated by the maximal mixing in the
stop sector \cite{Djouadi}. We have fixed the top quark mass to be 173.2 GeV.
We have not changed the above parameters, since they do not
affect the neutrino masses in the BRPVS model.
Indeed, for the above set of parameters in the squark sector, we have
almost got $m_h\sim$ 125 GeV, by varying the other parameters in the model.
As explained before, the neutrino masses in the BRPVS model are determined
by neutralino parameters ($M_1$, $M_2$, $\mu$, $\tan\beta$) and by
the masses of neutral Higgs bosons and sneutrinos.
We have chosen $M_1=\frac{5}{3}\tan^2\theta_W M_2$ and we have varied
$M_2$ from 100 GeV
to 1 TeV in steps of 100 GeV. As mentioned before that in order to have
$R_{\gamma\gamma}>1$, the mixing in the stau sector should be large, which
is determined by $\mu$, $\tan\beta$ and $(A_E)_{33}$. In our analysis,
we have varied $\mu$ from 100 GeV to 2 TeV in steps of 50 GeV and
$\tan\beta$ has been varied
from 5 to 60 in steps of 5. We have fixed $(A_E)_{33}=0$.
As explained before, while
solving for the neutrino mass eigenvalues, we have assumed degenerate masses
for sneutrinos. This would imply
that $(m_L^2)_{ij}=m_L^2\delta_{ij}$. For right-handed slepton masses,
we have assumed $(m_E^2)_{ij}=m_E^2\delta_{ij}$ and we have fixed $m_L^2=m_E^2$.
We vary the parameters $m_L$ and $m_A$,
which determine the sneutrino mass as well as the masses of CP-odd and heavy
Higgs bosons.

In the previous paragraph we have specified parameters of the model in order
to compute $m_h$ and $R_{\gamma\gamma}$. In fact, for fixed values of
$m_L$ and $m_A$, we scan over the neutralino parameters. In the scanning
procedure, we have demanded that
some constraints need to be satisfied. Among these, we have applied constraint
from the muon anomalous magnetic moment, $(g-2)_\mu$ \cite{mug-2}. The current
world average value of $(g-2)_\mu$ differs from its corresponding SM value
by about $3\sigma$ \cite{mug-2}. This discrepancy in $(g-2)_\mu$ is quantified
by $\Delta a_\mu$.
In the MSSM, at 1-loop level, $\Delta a_\mu$ gets contribution from
neutralino-charged slepton and chargino-sneutrino loops \cite{MSSMg-2}.
Since we have justified before that in the BRPVS model the additional
parameters $\epsilon_i$ and $(b_\epsilon)_i$ are very small, hence the
contribution to $\Delta a_\mu$ in the BRPVS model is almost same as
that in the MSSM. As a result of this, we have used the above mentioned
loop contributions to $\Delta a_\mu$ \cite{MSSMg-2} in our
numerical analysis. Below we describe the four constraints which we
have applied in our scanning procedure.
\begin{enumerate}[(i)]
\item $m_h$ should be in the range of 123 to 127 GeV,
\item either $R_{\gamma\gamma}>1$ or $R_{\gamma\gamma}\leq 1$,
\item masses of sleptons should be greater than 100 GeV,
\item $\Delta a_\mu$ should be in the range of $(1.1-4.7)\times 10^{-9}$.
\end{enumerate}
The constraints (i), (iii) and (iv) have been applied in every case.
Regarding the constraint (ii), we will specifically mention below
whether $R_{\gamma\gamma}>1$ or $R_{\gamma\gamma}\leq 1$ has been applied.
For those points in the parameter space which satisfy the above four
constraints, we calculate $\epsilon$ and $b_\epsilon$ which determine
the neutrino masses through Eq. (\ref{E:leadsol}). We have chosen the
following values for neutrino mass eigenvalues in order to be
consistent with the neutrino oscillation data \cite{osci}:
\begin{equation}
m_1=0,\quad m_2=\sqrt{\Delta m_{\rm sol}^2},\quad
m_3=\sqrt{\Delta m_{\rm atm}^2}.
\end{equation}
Here, the solar and atmospheric mass scales (central values), from a
global fit to neutrino oscillation data \cite{glob-fit}, are given
as $\Delta m_{\rm sol}^2=7.62\times 10^{-5}~{\rm eV}^2$ and
$\Delta m_{\rm atm}^2=2.55\times 10^{-3}~{\rm eV}^2$, respectively.

In Fig. 1 we have shown allowed values of $\epsilon$ and $b_\epsilon$
for different values of $m_L$ and $m_A$.
\begin{figure}[!h]
\begin{center}
\includegraphics[height=2.5in,width=2.5in]{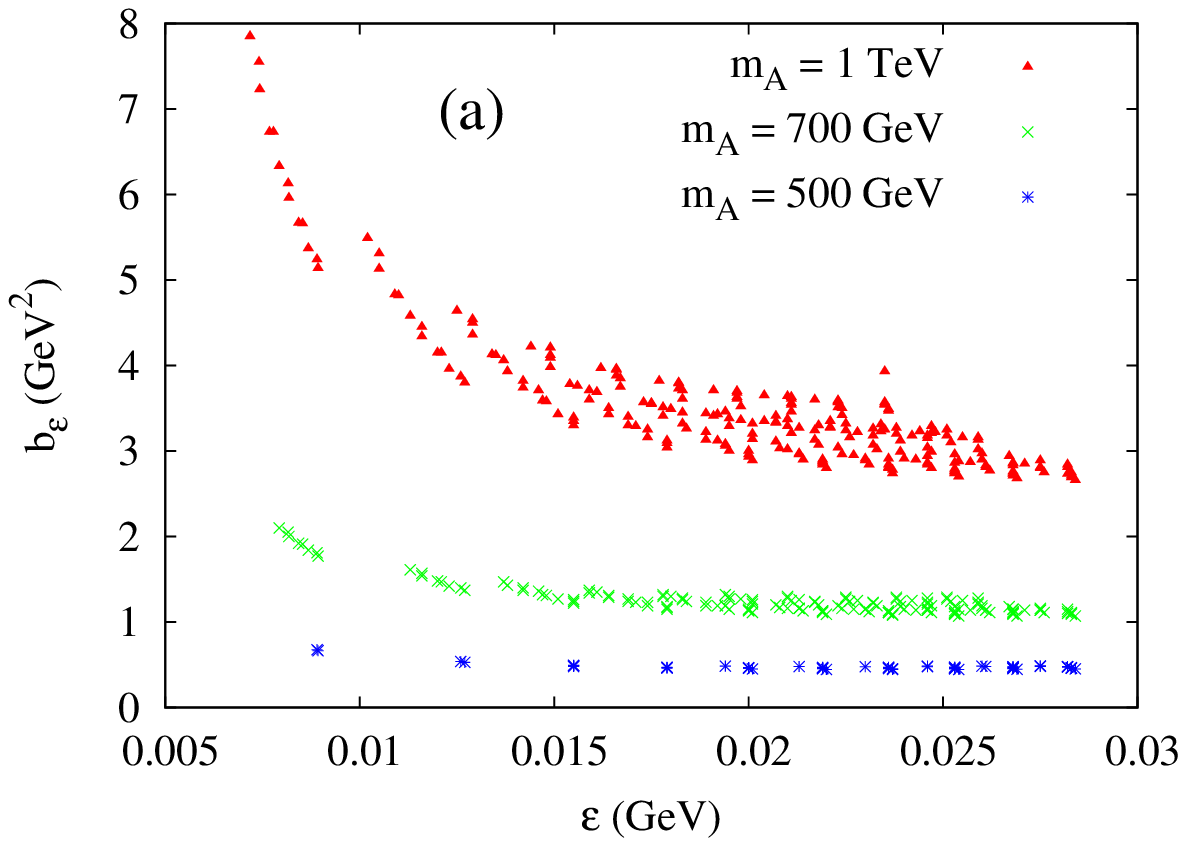}
\includegraphics[height=2.5in,width=2.5in]{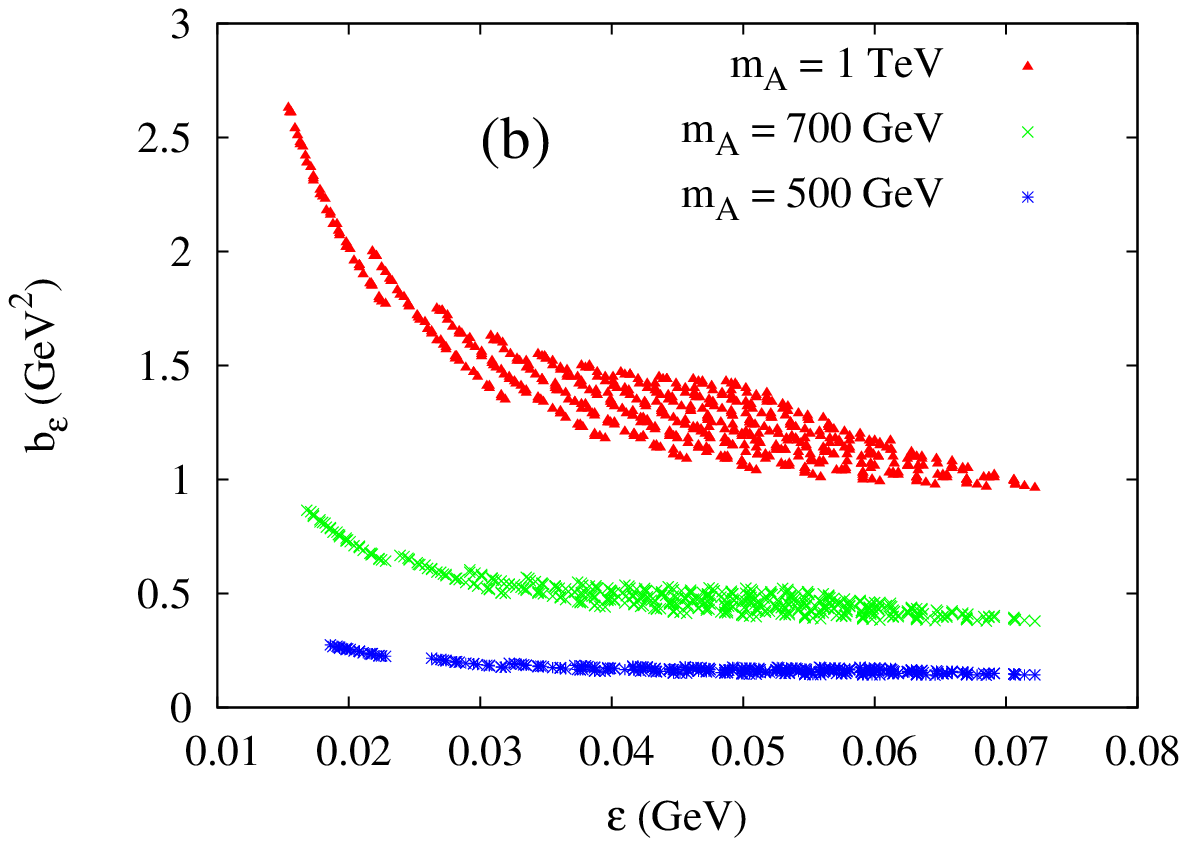}

\includegraphics[height=2.5in,width=2.5in]{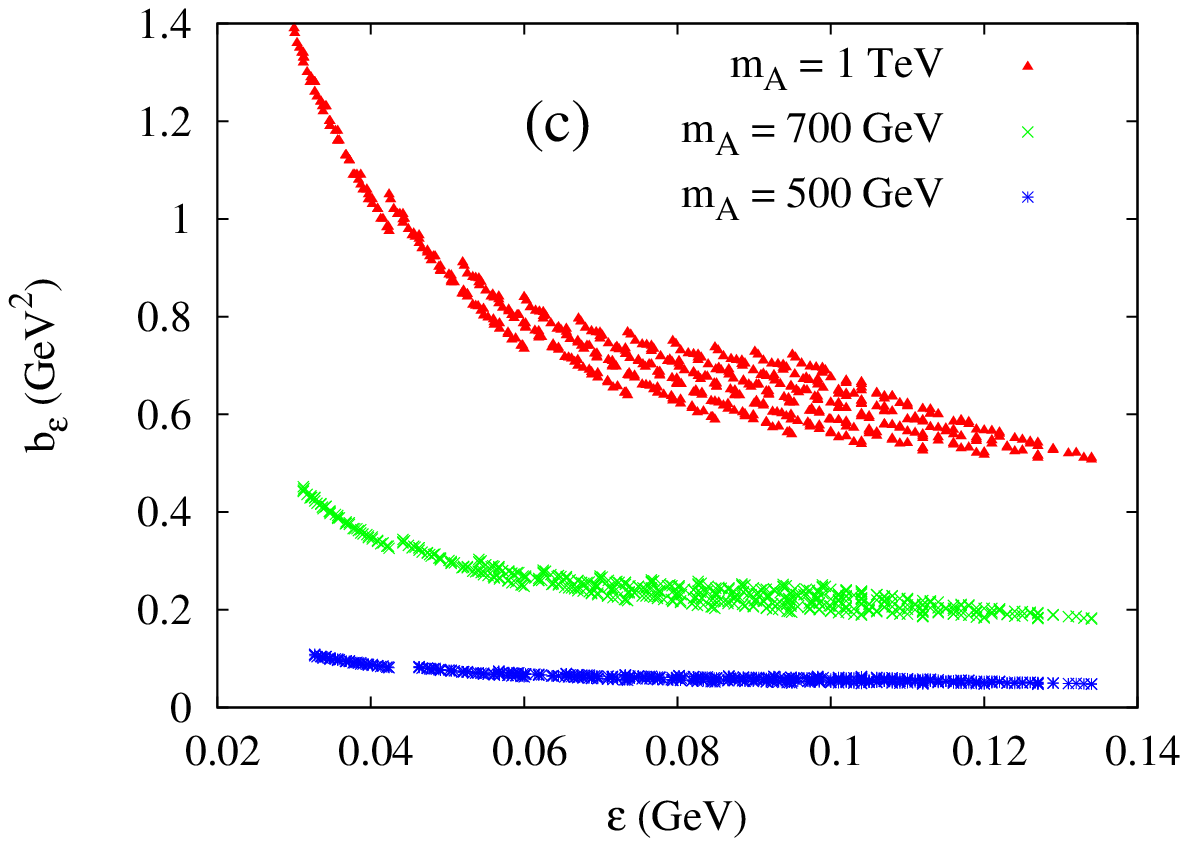}
\includegraphics[height=2.5in,width=2.5in]{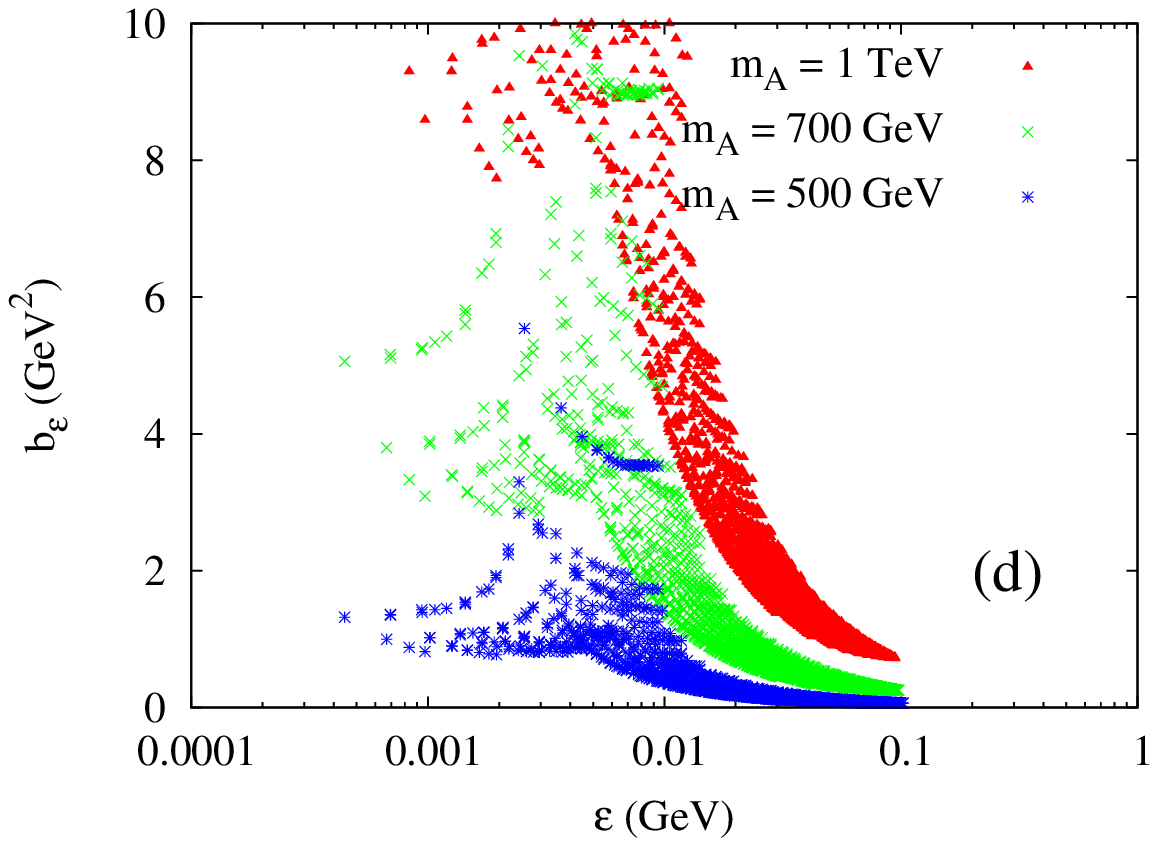}
\end{center}
\vspace*{-1cm}
\caption{Allowed values of $\epsilon$ and $b_\epsilon$. In plots (a), (b)
and (c) we have fixed $m_L=$ 200, 300 and 400 GeV, respectively,
and we have applied $R_{\gamma\gamma}>1$. In plot (d), $R_{\gamma\gamma}\leq 1$
has been applied and $m_L=$ 400 GeV. In each
of these plots red, green and blue color points represent $m_A=$
1000, 700 and 500 GeV, respectively.}
\end{figure}
In Fig. 1(d) we have applied the constraint $R_{\gamma\gamma}\leq 1$,
while in other plots of Fig. 1 the constraint
$R_{\gamma\gamma}>1$ has been applied.
In Fig. 1(a) $m_L$ has been kept to a very low value of 200 GeV, and in Figs.
1(b) and 1(c) $m_L=$ 300 and 400 GeV, respectively. As explained earlier,
the allowed points in these plots are satisfied by the requirement
$R_{\gamma\gamma}>1$. We have found that for $m_L=$ 150 GeV, the
constraint $R_{\gamma\gamma}>1$ is not satisfied. From the plots of
Fig. 1(a)$-$(c),
we can notice that the most likely value of $\epsilon$ is $\gapprox$ 0.01 GeV.
This value of $\epsilon$ is at least one order larger than its expected
value from the neutrino masses, which is described at the end of Sec. 2.1.
For $R_{\gamma\gamma}>1$, the lowest value of $\epsilon$ can
be found in Fig. 1(a), which is $\approx$ 0.007 GeV, and at these points
$M_2$ should be as low as 100 GeV. On the other hand, in future, if LHC has not
found any excess in the Higgs to diphoton decay rate, then from Fig. 4(d)
we can notice that we can satisfy the neutrino oscillation data for
$\epsilon$ between about $10^{-4}$ to 0.1 GeV. In Fig. 4(d) we have
fixed $m_L=$ 400 GeV. By decreasing $m_L$, the allowed space for $\epsilon$
and $b_\epsilon$ is slightly different from that of Fig. 4(d).
Hence, from the measurement of Higgs to diphoton decay rate,
if $R_{\gamma\gamma}>1$ then $\epsilon$ should
be at least $\sim 10^{-2}$ GeV. Otherwise, if $R_{\gamma\gamma}\leq 1$
then $\epsilon$ can be as low as $\sim 10^{-4}$ GeV.

From the plots of Fig. 1(a)$-$(c), we can notice
that when we increase $m_L$ by keeping $m_A$ fixed, the average value of
$\epsilon$ is increasing. By increasing $m_L$, the lower limit on
$\tan\beta$ and $\mu$ will increase in the case of $R_{\gamma\gamma}>1$,
which we will describe below.
As a result of this, the quantity $a_0$ of Eq. (\ref{E:a0a1}), which
is inversely related to $\epsilon$, will decrease.

In Sec. 2.1, from the neutrino mass scale we have estimated that
$b_\epsilon$ is $\sim$ 1 GeV$^2$. In the case of
$R_{\gamma\gamma}>1$, from the plots of Fig. 1(a)$-$(c), we can notice that for
$m_A\gapprox$ 700 GeV, $b_\epsilon$ can be in the range of
0.5 to 2 GeV$^2$. For $R_{\gamma\gamma}>1$, the lowest value of
$b_\epsilon$ has been found to be
about 0.05 GeV$^2$, which can be seen in Fig. 1(c). In Fig. 1(d) we have
$R_{\gamma\gamma}\leq 1$, and the allowed value of $b_\epsilon$ can
be around 1 GeV$^2$ by appropriately choosing the $m_A$. For instance,
if we have to achieve $\epsilon\sim 10^{-3}$ GeV and $b_\epsilon\sim$
1 GeV$^2$, then $m_A$ should be $\lapprox$ 500 GeV.

In Figs. 1(a)$-$(c), for a given value of
$m_L$, $b_\epsilon$ is increasing with $m_A$. The reason for this is that
$b_\epsilon$ is inversely related to $a_1$, and from Eq. (\ref{E:a0a1})
we can understand that $a_1$ decreases with $m_A$. Similarly,
from Figs. 1(a)$-$(c),
by keeping $m_A$ fixed, we can notice that the average value of $b_\epsilon$
is decreasing with $m_L$. We will shortly explain below that by increasing
$m_L$ the lower limit on $\tan\beta$ and $\mu$ will increase in the
case of $R_{\gamma\gamma}>1$. Hence,
although the function $I_4$ of $a_1$ decreases with increasing $m_L$, the
factor $\frac{1}{\cos^2\beta}$ in $a_1$ will compensate this decrease,
and the net result is that $a_1$ increases with $m_L$.

In Fig. 2 we have plotted allowed values of $\mu$ and $\tan\beta$.
The points in Fig. 2(a) are allowed by the constraint $R_{\gamma\gamma}> 1$,
whereas the points in Fig. 2(b) satisfy $R_{\gamma\gamma}\leq 1$.
In the plots of Fig. 2 there are no allowed points for $\tan\beta = 5$.
We have found that for such a low $\tan\beta$ the mass of light Higgs boson is
below 123 GeV and hence do not satisfy the constraint (i).
\begin{figure}[!h]
\begin{center}
\includegraphics[height=2.5in,width=2.5in]{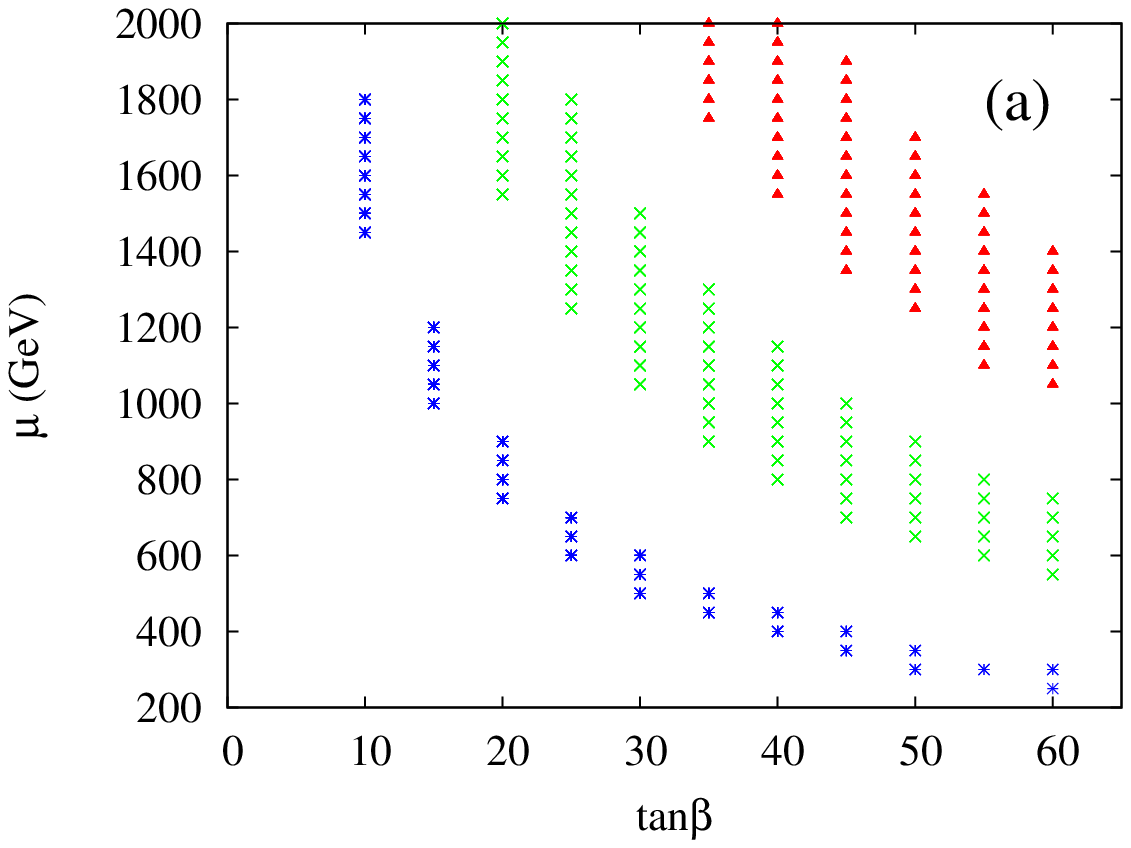}
\includegraphics[height=2.5in,width=2.5in]{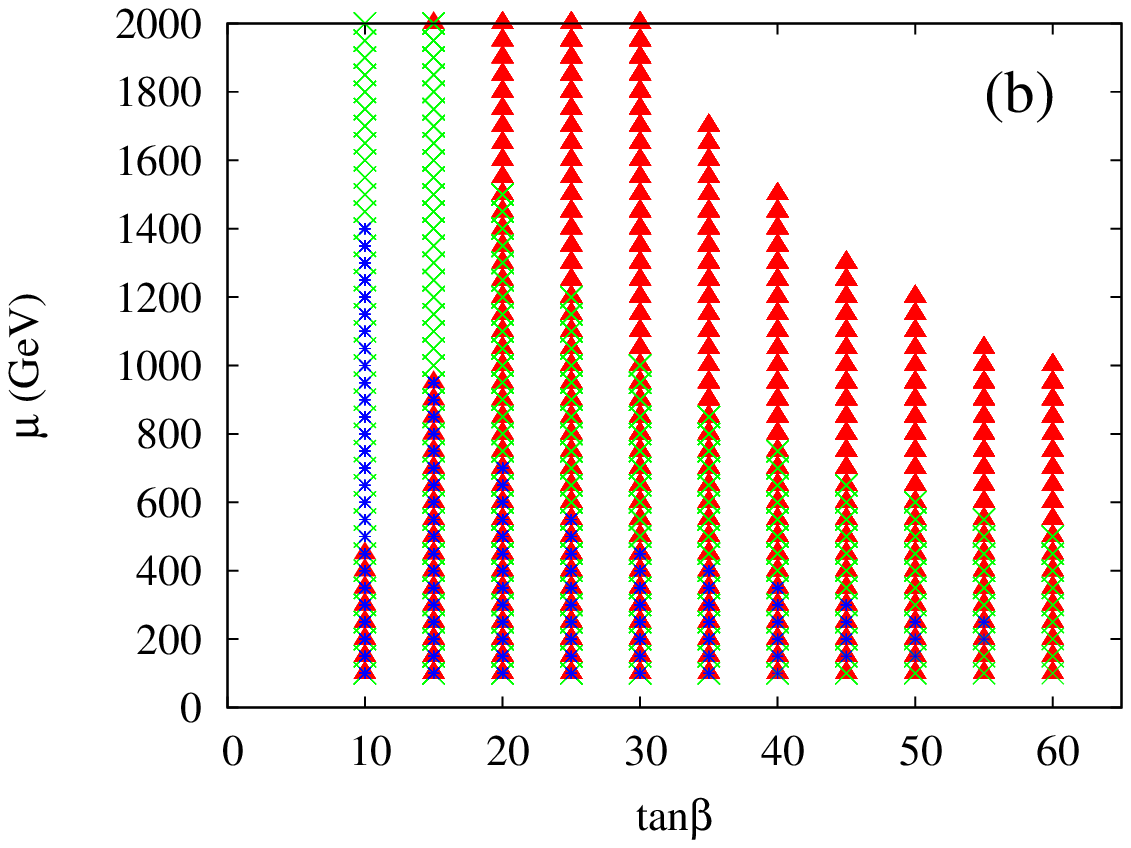}
\end{center}
\caption{Allowed values of $\mu$ and $\tan\beta$ for $m_A$ = 1 TeV.
In plots (a) and (b) we have applied the constraints $R_{\gamma\gamma}>1$
and $R_{\gamma\gamma}\leq 1$, respectively.
The red, green and blue points are for $m_L$ = 400, 300 and 200 GeV,
respectively.}
\end{figure}
In Fig. 2(a), we can notice that by increasing the value of $m_L$,
the lower limit on $\mu$ and $\tan\beta$ would increase. We may
understand this from the fact that the stau masses should be as light
as possible, and moreover, the mixing in the stau sector should be large in
order to have $R_{\gamma\gamma}>1$ \cite{Carena-etal}.
Hence, by increasing the soft mass $m_L$, the quantity $\mu\times\tan\beta$
should proportionately be increased in order to decrease the lightest
stau mass and also to increase the mixing in the stau sector. From Fig. 2(a),
we can notice that for a specific value of $\tan\beta$ the allowed
value of $\mu$ lies in a certain range. We have seen that the lower and
upper limits of this range are restricted by the constraints
(ii) and (iii). For instance, for $m_L$ =
300 GeV and $\tan\beta$ = 30, the allowed range for $\mu$ is
from 1050 to 1500 GeV. In this case, for $\mu <$ 1050 GeV we
may not satisfy $R_{\gamma\gamma}>1$. On the other hand, for $\mu >$ 1500 GeV
the lightest stau mass becomes less than 100 GeV. In Fig. 2(a) we have
fixed $m_A$ = 1 TeV. By decreasing $m_A$, we have found that $\tan\beta$ is not
restricted, however, for each $\tan\beta$ the corresponding lower limit
on $\mu$ will increase. To illustrate this point, by considering the case of
$m_A$ = 700 GeV, $m_L$ = 300 GeV and $\tan\beta$ = 30, the allowed
range for $\mu$ has been found to be between 1150 to 1500 GeV.
Hence, these results
indicate that by decreasing $m_A$ the $R_{\gamma\gamma}$ value will
decrease.

As stated before, in Fig. 2(b) we have applied the constraint
$R_{\gamma\gamma}\leq 1$. In this plot we can
see that $\mu$ can be as low as 100 GeV.
Numerically, we have noticed that $R_{\gamma\gamma}$ increases with $\mu$
and hence after a certain large value of $\mu$, $R_{\gamma\gamma}\leq 1$
may not
be satisfied. In the case of $m_L=$ 400 GeV, in Fig. 2(b), for $\tan\beta=$
10 and 15, large values of $\mu$ are not allowed by the constraint (iv).
In fact,
allowed points in Fig. 2(b) indicate that $R_{\gamma\gamma}\leq 1$ can
be satisfied for large $\tan\beta$ and relatively large $\mu$ values.
For these large values of $\mu$ and $\tan\beta$, the calculated values
of $\epsilon$ can be as high as 0.1 GeV, which can be seen in Fig. 1(d).
For low $\tan\beta$ and moderate values of $\mu$, $\epsilon$ can
be $\lapprox 10^{-3}$ GeV.

In Fig. 3, we show the correlation between enhancement in the
Higgs to diphoton decay rate ($R_{\gamma\gamma}$) and the bilinear parameter
$\epsilon$.
\begin{figure}
\begin{center}
\includegraphics[height=2.5in,width=2.5in]{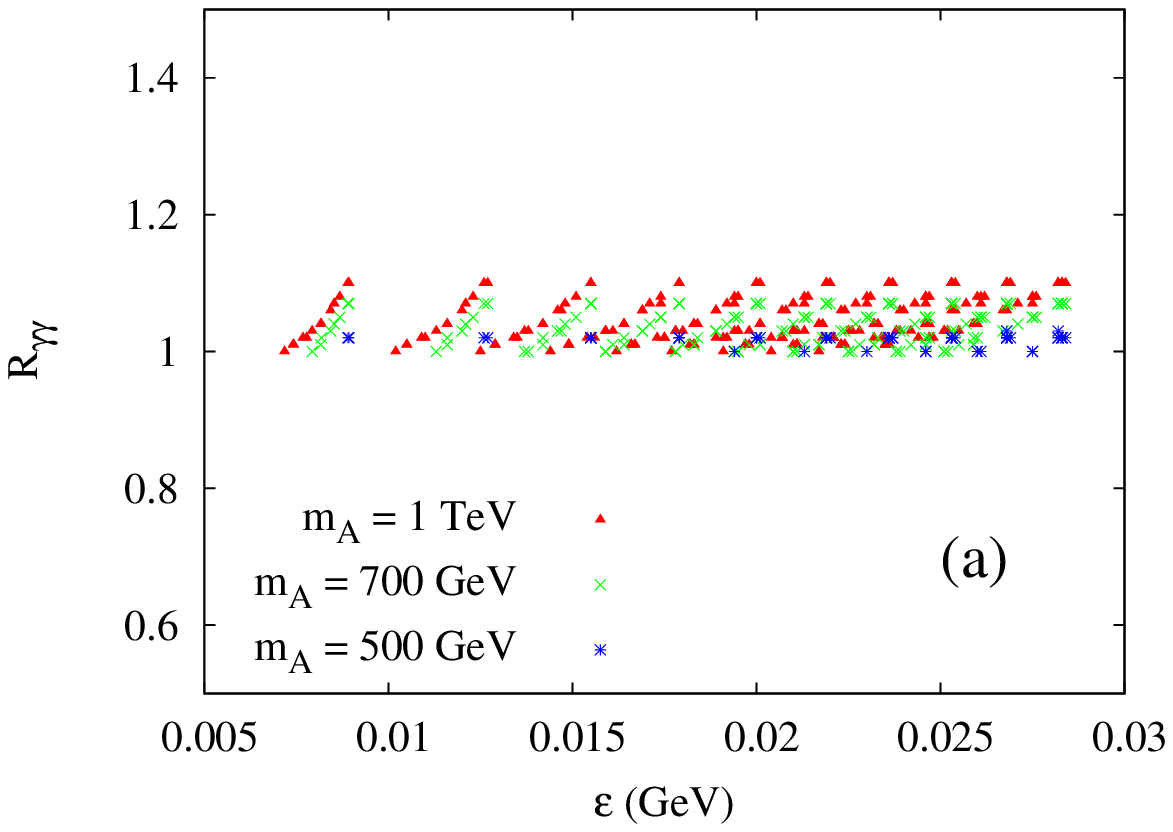}
\includegraphics[height=2.5in,width=2.5in]{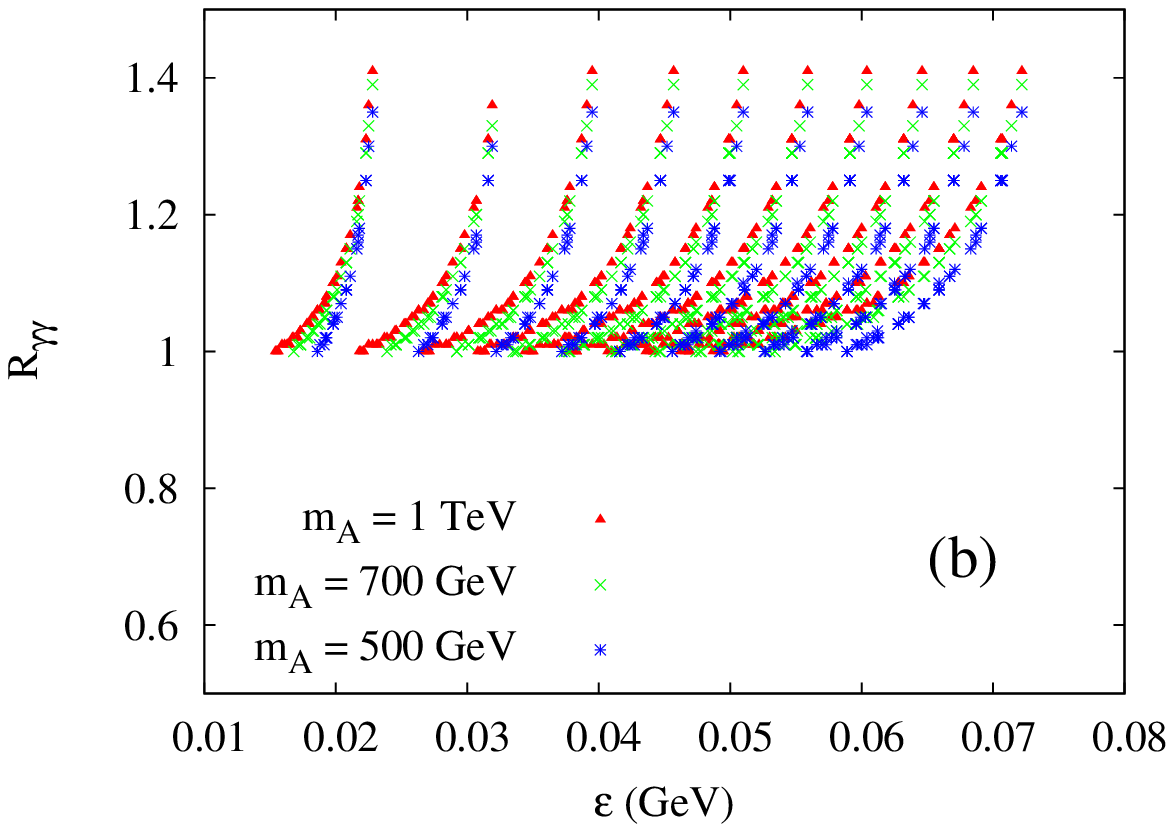}

\includegraphics[height=2.5in,width=2.5in]{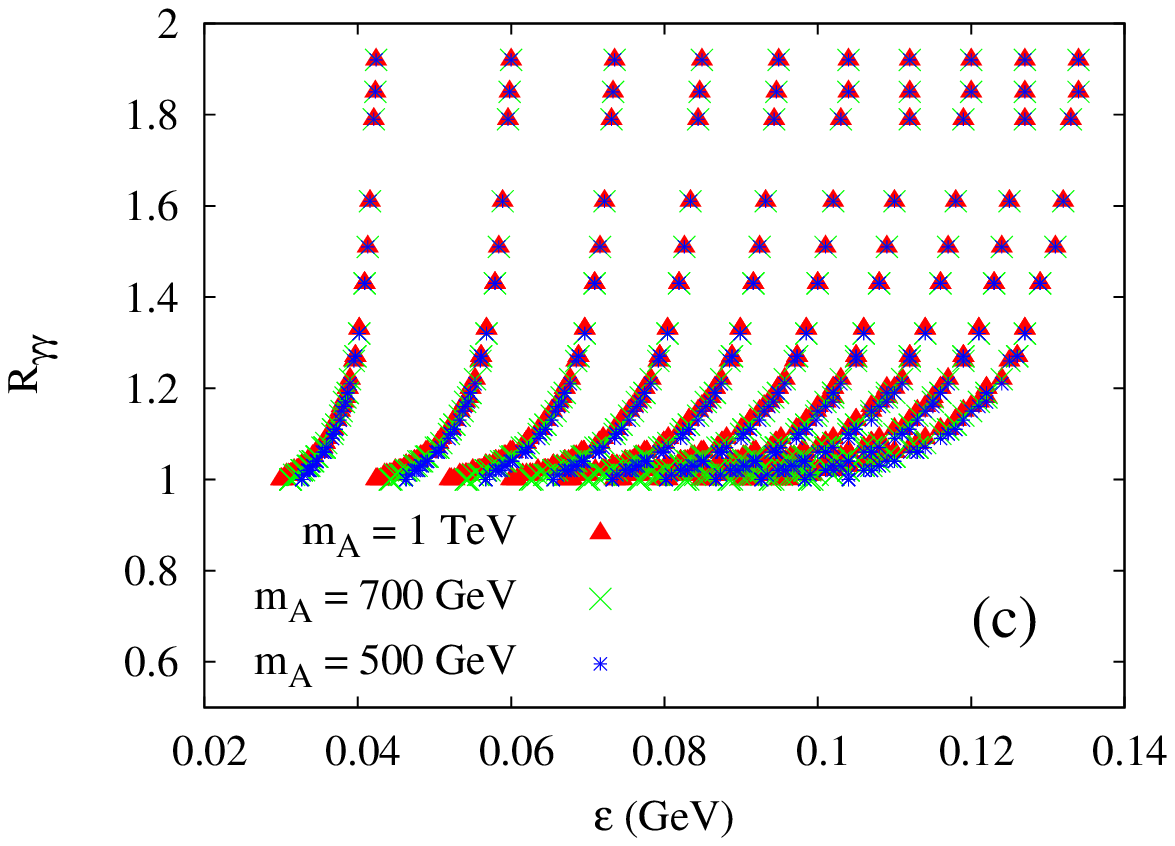}
\includegraphics[height=2.5in,width=2.5in]{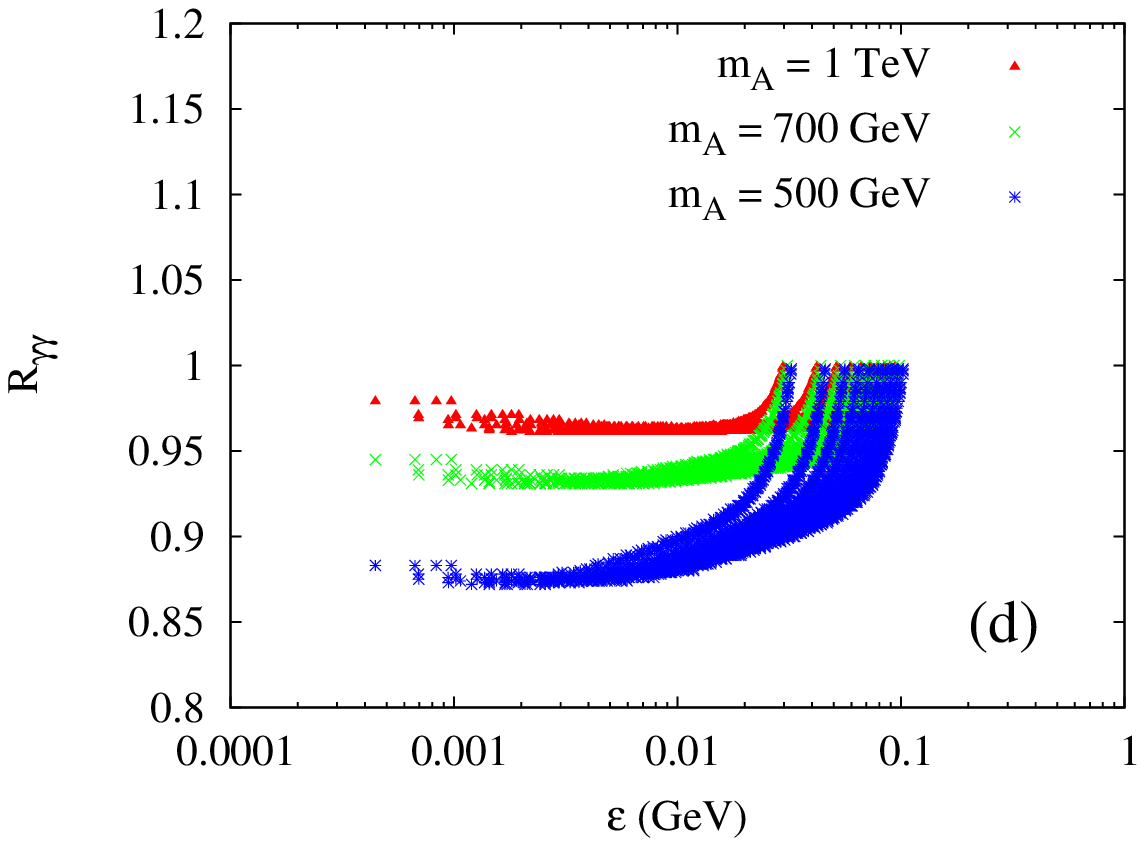}
\end{center}
\caption{Allowed values of $\epsilon$ versus $R_{\gamma\gamma}$. In the
plots (a), (b) and (c), the value of $m_L$ = 200, 300 and 400 GeV,
respectively. In these plots we have applied the constraint
$R_{\gamma\gamma}>1$. In plot (d), $m_L=$ 400 GeV and the constraint
$R_{\gamma\gamma}\leq 1$ has been applied. In each
of these plots red, green and blue color points represent $m_A=$
1000, 700 and 500 GeV, respectively.}
\end{figure}
From Figs. 3(a)$-$(c), we can observe that for a low value of $m_L$ = 200 GeV,
the maximum value for $R_{\gamma\gamma}$ is $\approx$ 1.1. As noted before,
in the case of $R_{\gamma\gamma}>1$, the lowest value of $\epsilon$ is
$\approx$ 0.007 GeV, which is found for $m_L=$ 200 GeV.
For this lowest value of $\epsilon$ the $R_{\gamma\gamma}$ value is barely
greater than 1.0. From Figs. 3(a)$-$(c), we can notice that the
maximum value of
$R_{\gamma\gamma}$ is increasing with $m_L$. We have stated before that
by increasing $m_L$, the values of $\mu$ and $\tan\beta$ would increase.
For large values of $\mu$ and $\tan\beta$, the coupling strengths of
staus to the light Higgs boson would increase \cite{Carena-etal}.
As a result of this, $R_{\gamma\gamma}$ is increasing with $m_L$.
The maximum values of $R_{\gamma\gamma}$
in Figs. 3(b) and 3(c) are 1.41 and 1.92 respectively.
We may increase
$R_{\gamma\gamma}$ to more than 2.0 by increasing $m_L$ to 500 GeV. But
in order to have large mixing and lower masses in the stau sector, we
have to proportionately increase $\mu$ and $\tan\beta$. In this
work we have scanned $\mu$ and $\tan\beta$ up to 2 TeV and 60, respectively,
and have not considered cases of $m_L\geq$ 500 GeV. However, from the
above mentioned arguments, for $R_{\gamma\gamma}>1$, we can speculate
that by increasing $m_L$ to 500 GeV the value of $\epsilon$ would be
around 0.1 GeV.

In Fig. 3(d), we have applied the constraint $R_{\gamma\gamma}\leq 1$.
We can notice from this plot that for $\epsilon\sim 10^{-3}$ GeV,
$R_{\gamma\gamma}$ is different for different values of $m_A$. From this
perspective we can argue that, if $R_{\gamma\gamma}\leq 1$ is found
to be true, then a precise measurement of $R_{\gamma\gamma}$ can be
used to determine $\epsilon$ and $m_A$.

\subsection{Smallness of $\epsilon$ and $b_\epsilon$}

In this subsection, we try to motivate the smallness of $\epsilon$
and $b_\epsilon$ from a high scale physics. Essentially we will explore
what the Higgs to diphoton decay rate can tell us about the high scale
physics parameters. As it is noted in \cite{HPT},
by assuming supersymmetry breaking at an intermediate energy scale
$\Lambda\sim 10^{11}$ GeV, we can explain the $\mu$-parameter, soft terms
in scalar potential
as well as $\epsilon$- and $b_\epsilon$- terms.
Here we briefly describe important
ingredients from Ref. \cite{HPT}. By introducing SM gauge singlet
superfields $\hat{S}$, $\hat{X}_1$ and $\hat{X}_2$, we may write the
superpotential as
\begin{equation}
W=\Lambda^2\hat{S}+\frac{1}{M_P}\hat{X}_1^3\hat{X}_2+\frac{\hat{X}_1^2}{M_P}
\hat{H}_u\hat{H}_d+\frac{\hat{X}_2^3}{M^2_P}\hat{L}\hat{H}_u+\cdots.
\label{E:W}
\end{equation}
Here, $M_P=2.4\times 10^{18}$ GeV is the Planck scale. There can
be ${\cal O}(1)$ couplings in the above terms, which we have neglected.
In the above equation we have written only the necessary terms for our
purpose, and these terms can be justified by introducing additional
symmetries, say gauged $U(1)^\prime$. $\hat{S}$ must be singlet under this
additional $U(1)^\prime$, but $\hat{X}_{1,2}$ can be charged under
$U(1)^\prime$. The vevs of the scalar components of these SM gauge
singlet superfields can be arranged as \cite{HPT}: $\langle S\rangle\sim M_P$,
$\langle X_{1,2}\rangle=\Lambda_{1,2}\sim\Lambda$.
The first term of Eq. (\ref{E:W})
breaks supersymmetry spontaneously by acquiring an auxiliary vev
to $\hat{S}$ which is of the
order of $\Lambda^2$. This axillary vev can generate soft
terms in the scalar potential with mass parameters $m_{\rm soft}\sim
\frac{\Lambda^2}{M_P}$. Here, the generation of soft terms in the scalar
potential is based on the Polonyi mechanism \cite{Polonyi}.
The scalar vevs of $\hat{X}_{1,2}$ generate the
$\mu$- and $\epsilon$-parameters which are $\mu\sim \frac{\Lambda_1^2}{M_P}$
and $\epsilon\sim\frac{\Lambda_2^3}{M_P^2}$. Here, we have followed the
Kim-Nilles mechanism for the generation of $\mu$-term \cite{Kim-Nilles}.
The auxiliary
vevs of $\hat{X}_{1,2}$ can generate $b_\mu$ and $b_\epsilon$ which are
$b_\mu\sim\frac{\Lambda_1^3\Lambda_2}{M_P^2}$ and $b_\epsilon\sim
\frac{\Lambda_1^3\Lambda_2^2}{M_P^3}$.

In the previous paragraph, we have given motivation for the generation
of $\epsilon$
and $b_\epsilon$ parameters as well as other supersymmetric parameters
from a high scale physics. Now we have to fix the high scale physics
parameters in order to fit the low energy data.
Since we expect $m_{\rm soft}\sim\mu\sim$ TeV, for $\Lambda\sim\Lambda_1
\sim 0.5\times 10^{11}$ GeV we can explain the TeV scale masses
for supersymmetric
fields. If we take $\Lambda_2\sim 10^{11}$ GeV we can get
$\epsilon\sim 10^{-3}$ GeV. From Figs. 1 and 3, we can say that a value
of $\epsilon\sim 10^{-3}$ GeV is consistent with $R_{\gamma\gamma}\leq 1$.
In order to achieve $R_{\gamma\gamma}>1$, $\epsilon$ should be $\gapprox$
0.01 GeV. Hence, by taking $\Lambda_2\sim 3.9\times 10^{11}$ GeV we can get
$\epsilon\sim 10^{-2}$ GeV. So, if $R_{\gamma\gamma}>1$ then there is
a little hierarchy between $\Lambda_1$ and $\Lambda_2$, otherwise, this
hierarchy can be reduced.

In future, if LHC has found that $R_{\gamma\gamma}$
is significantly larger than 1.0, then from Figs. 3(a)$-$(c) we can
say that $m_L$ should be larger than about 300 GeV. For $m_L$ between
300 to 400 GeV, from Figs. 1(b)and 1(c) we can have $b_\epsilon\sim$
1 GeV$^2$, provided $m_A$ is $\gapprox$ 700 GeV. Now, for $\Lambda_1\sim
0.5\times 10^{11}$ GeV and $\Lambda_2\sim 3.9\times 10^{11}$ GeV, we
can get $b_\mu\sim$ 8 TeV$^2$ and $b_\epsilon\sim$ 1 GeV$^2$. Hence,
for the case of $R_{\gamma\gamma}>1$, we can motivate consistent
supersymmetric mass spectrum and 0.1 eV scale for neutrino masses from a high
scale physics by proposing two different intermediate scales. The
hierarchy between these two scales should be at least 8.

If there is no enhancement in the Higgs to diphoton decay rate, from
Fig. 1(d), we can notice that $\epsilon$ can be between $\sim 10^{-3}$ to
0.1 GeV. From the above discussion, to achieve $\epsilon\sim$ 0.1 GeV
from high scale physics, there need to be little hierarchy between the
intermediate scales $\Lambda_{1,2}$. This hierarchy can be minimal for
$\epsilon\sim 10^{-3}$ GeV. For $\epsilon\sim 10^{-3}$ GeV, in Fig. 1(d),
$b_\epsilon$ can be around 1 GeV$^2$ if $m_A\sim$ 500 GeV. For
this set of values, from Fig. 3(d), we can notice that $R_{\gamma\gamma}$
is little less than 0.9. Hence, if we believe in the motivation
of $\epsilon$ and $b_\epsilon$ from high scale physics, the high energy
scales in this scenario depend on the value of $R_{\gamma\gamma}$.
Moreover, in the above described analysis, we can also estimate $m_L$
and $m_A$ by knowing the $R_{\gamma\gamma}$.
So the future runs at the LHC can give us clue about this scenario
by precisely finding the $R_{\gamma\gamma}$. 

We make short comments about measuring the parameters $\epsilon$ and
$b_\epsilon$ in experiments. Since both these parameters indicate
that R-parity is violated, a consequence of that is that the lightest
supersymmetric particle (LSP) is unstable. Depending on the parameter
space, either the lightest neutralino or the lightest charged slepton
can be the LSP in this model \cite{TestRp}. The decay life time of LSP
is determined
by $\epsilon$ and $b_\epsilon$. Hence the signals of the decay of LSP
in this model should give an indication about these bilinear
parameters \cite{TestRp}, from which we can verify the neutrino mass
mechanism and also its correlation to the Higgs to diphoton decay rate.

\section{Conclusions}

Recently the LHC has discovered a Higgs-like particle whose mass
is around 125 GeV. It has also been indicated that there is an enhancement
in the Higgs to diphoton decay rate as compared to that in the SM. We have
studied implications of these discoveries in the BRPVS model. This
model is a minimal extension of the MSSM where the bilinear terms
$\epsilon\hat{L}\hat{H}_u$ and $b_\epsilon\tilde{L}H_u$ are added to
the superpotential and scalar potential of the model, respectively.
The main objective
of this model is to explain the smallness of neutrino masses, where
the neutrino mass eigenvalues can be shown to be dependent on neutralino
parameters, soft mass for charged sleptons ($m_L$) and CP-odd Higgs
boson mass ($m_A$)
\cite{Davidson-etal,GR}, apart from the bilinear parameters $\epsilon$
and $b_\epsilon$.

In our analysis, we have scanned over the neutralino parameters and varied
$m_L$ and $m_A$ accordingly. We have also fixed the soft masses for
third generation squarks, in order to have light Higgs boson mass to
be around 125 GeV. We then have studied implications of enhancement in the
Higgs to diphoton decay rate ($R_{\gamma\gamma}$) in the BRPVS model.
Explicitly we have found that in order to
be compatible with $R_{\gamma\gamma}>1$ and the neutrino oscillation data,
the unknown bilinear parameter should be $\epsilon\gapprox 10^{-2}$ GeV.
We have also found that to achieve
$R_{\gamma\gamma}$ between about 1.5 to 2.0, $m_L$ should be between
300 to 400 GeV, provided $\mu$ and $\tan\beta$ are scanned up to 2 TeV
and 60 respectively. We have not obtained specific bounds on $m_A$.
However, from the order of estimations we expect $b_\epsilon$ to be around
1 GeV$^2$ and to achieve this with the above mentioned $m_L$, $m_A$
can be in the range of 700 to 1000 GeV.

Since $R_{\gamma\gamma}>1$ is not yet confirmed by LHC, we have also
analyzed the case of $R_{\gamma\gamma}\leq 1$. In this later case, we
have found that $\epsilon$ can be between $\sim 10^{-3}$ to 0.1 GeV. The
corresponding $b_\epsilon$ can be around 1 GeV$^2$ by appropriately
choosing $m_A$ to be from 500 to 1000 GeV. Moreover, we have also
found that $R_{\gamma\gamma}$ can be as low as 0.85.

From the above discussion, we can notice that $\epsilon$ and $b_\epsilon$
need to be very small in GeV units. We have motivated smallness
of these two parameters from a high scale physics, and at the same
time we have also explained the TeV scale masses for supersymmetric fields.
We have found that to explain $\epsilon\sim 10^{-2}$ GeV and $b_\epsilon
\sim$ 1 GeV$^2$ there need to be two different intermediate
scales ($\sim 10^{11}$ GeV)  with a hierarchy of factor of 8 between them.

\section*{Acknowledgments}

The author is thankful to Sudhir Vempati for discussions and also for
reading the manuscript. The author acknowledges techincal help from
Debtosh Chowdhury.


\begin{thebibliography}{99}
\bibitem{LHC}
  G.~Aad {\it et al.}  [ATLAS Collaboration],
    Phys.\ Lett.\ B {\bf 716}, 1 (2012)
    [arXiv:1207.7214 [hep-ex]];
  S.~Chatrchyan {\it et al.}  [CMS Collaboration],
    Phys.\ Lett.\ B {\bf 716}, 30 (2012)
    [arXiv:1207.7235 [hep-ex]].

\bibitem{Higgs}
  F.~Englert and R.~Brout,
  Phys.\ Rev.\ Lett.\  {\bf 13}, 321 (1964);
  P.~W.~Higgs,
  Phys.\ Lett.\  {\bf 12}, 132 (1964);
  P.~W.~Higgs,
  Phys.\ Rev.\ Lett.\  {\bf 13}, 508 (1964);
  G.~S.~Guralnik, C.~R.~Hagen and T.~W.~B.~Kibble,
  Phys.\ Rev.\ Lett.\  {\bf 13}, 585 (1964);
  P.~W.~Higgs,
  Phys.\ Rev.\  {\bf 145}, 1156 (1966);
  T.~W.~B.~Kibble,
  Phys.\ Rev.\  {\bf 155}, 1554 (1967).

\bibitem{Hubaut}
Please see the talk by F. Hubaut in the conference Rencontres de Moriond
(EW Interactions and Unified Theories),
https://indico.in2p3.fr/conferenceDisplay.py?confId=7411

\bibitem{Ochando}
Please see the talk by C. Ochando in the conference Rencontres de Moriond
(QCD and High Energy Interactions), http://moriond.in2p3.fr/QCD/2013/qcd.html

\bibitem{susy}
  H.~P.~Nilles, Phys. Rept. {\bf 110}, 1 (1984);
  H.~E.~Haber and G.~L.~Kane, Phys. Rept. {\bf 117}, 75 (1985);
  S.~P.~Martin, arXiv:hep-ph/9709356;
  M.~Drees, R.~Godbole and P.~Roy, Theory and Phenomenology of Sparticles,
  (World Scientific, 2004);
  P.~Binetruy, Supersymmetry (Oxford University Press, 2006);
  H.~Baer and X.~Tata, Weak Scale Supersymmetry: From
  Superfields to Scattering Events, (Cambridge University Press, 2006).

\bibitem{osci}
  Y.~Fukuda {\it et al.}  [Kamiokande Collaboration],
  Phys.\ Rev.\ Lett.\  {\bf 77}, 1683 (1996);
  W.~Hampel {\it et al.}  [GALLEX Collaboration],
  Phys.\ Lett.\ B {\bf 447}, 127 (1999);
  J.~N.~Abdurashitov {\it et al.}  [SAGE Collaboration],
  Phys.\ Rev.\ C {\bf 60}, 055801 (1999)
  [astro-ph/9907113];
  Q.~R.~Ahmad {\it et al.}  [SNO Collaboration],
  Phys.\ Rev.\ Lett.\  {\bf 87}, 071301 (2001)
  [nucl-ex/0106015];
  K.~Eguchi {\it et al.}  [KamLAND Collaboration],
  Phys.\ Rev.\ Lett.\  {\bf 90}, 021802 (2003)
  [hep-ex/0212021];
  J.~Hosaka {\it et al.}  [Super-Kamiokande Collaboration],
  Phys.\ Rev.\ D {\bf 73}, 112001 (2006)
  [hep-ex/0508053];
  Y.~Fukuda {\it et al.}  [Super-Kamiokande Collaboration],
  Phys.\ Rev.\ Lett.\  {\bf 81}, 1562 (1998)
  [hep-ex/9807003];
  M.~Ambrosio {\it et al.}  [MACRO Collaboration],
  Phys.\ Lett.\ B {\bf 434}, 451 (1998)
  [hep-ex/9807005].

\bibitem{cosmo}
  E.~Komatsu {\it et al.}  [WMAP Collaboration],
  Astrophys.\ J.\ Suppl.\  {\bf 180}, 330 (2009)
  [arXiv:0803.0547 [astro-ph]];
  J.~Dunkley {\it et al.}  [WMAP Collaboration],
  Astrophys.\ J.\ Suppl.\  {\bf 180}, 306 (2009)
  [arXiv:0803.0586 [astro-ph]].

\bibitem{beta}
  C.~.Kraus, B.~Bornschein, L.~Bornschein, J.~Bonn, B.~Flatt, A.~Kovalik, B.~Ostrick and E.~W.~Otten {\it et al.},
  Eur.\ Phys.\ J.\ C {\bf 40}, 447 (2005)
  [hep-ex/0412056].

\bibitem{BRPVSmod}
  M.~Hirsch and J.~W.~F.~Valle,
  New J.\ Phys.\  {\bf 6}, 76 (2004)
  [hep-ph/0405015].

\bibitem{Hirsch-etal}
  M.~Hirsch, M.~A.~Diaz, W.~Porod, J.~C.~Romao and J.~W.~F.~Valle,
  Phys.\ Rev.\ D {\bf 62}, 113008 (2000)
  [Erratum-ibid.\ D {\bf 65}, 119901 (2002)]
  [hep-ph/0004115];
  M.~A.~Diaz, M.~Hirsch, W.~Porod, J.~C.~Romao and J.~W.~F.~Valle,
  Phys.\ Rev.\ D {\bf 68}, 013009 (2003)
  [Erratum-ibid.\ D {\bf 71}, 059904 (2005)]
  [hep-ph/0302021].

\bibitem{Davidson-etal}
  S.~Davidson and M.~Losada,
  JHEP {\bf 0005}, 021 (2000)
  [hep-ph/0005080],
  Phys.\ Rev.\ D {\bf 65}, 075025 (2002)
  [hep-ph/0010325].

\bibitem{GR}
  Y.~Grossman and S.~Rakshit,
  Phys.\ Rev.\ D {\bf 69}, 093002 (2004)
  [hep-ph/0311310].

\bibitem{Rpsmall}
  A.~Masiero and J.~W.~F.~Valle,
  Phys.\ Lett.\ B {\bf 251}, 273 (1990);
  J.~C.~Romao, C.~A.~Santos and J.~W.~F.~Valle,
  Phys.\ Lett.\ B {\bf 288}, 311 (1992);
  J.~C.~Romao, A.~Ioannisian and J.~W.~F.~Valle,
  Phys.\ Rev.\ D {\bf 55}, 427 (1997)
  [hep-ph/9607401].

\bibitem{HPT}
  R.~S.~Hundi, S.~Pakvasa and X.~Tata,
  Phys.\ Rev.\ D {\bf 79}, 095011 (2009)
  [arXiv:0903.1631 [hep-ph]].

\bibitem{Rpheno}
  A.~S.~Joshipura and M.~Nowakowski,
  Phys.\ Rev.\ D {\bf 51}, 2421 (1995)
  [hep-ph/9408224];
  R.~Hempfling,
  Nucl.\ Phys.\ B {\bf 478}, 3 (1996)
  [hep-ph/9511288];
  S.~Roy and B.~Mukhopadhyaya,
  Phys.\ Rev.\ D {\bf 55}, 7020 (1997)
  [hep-ph/9612447];
  B.~Mukhopadhyaya, S.~Roy and F.~Vissani,
  Phys.\ Lett.\ B {\bf 443}, 191 (1998)
  [hep-ph/9808265];
  S.~Y.~Choi, E.~J.~Chun, S.~K.~Kang and J.~S.~Lee,
  Phys.\ Rev.\ D {\bf 60}, 075002 (1999)
  [hep-ph/9903465];
  A.~S.~Joshipura, R.~D.~Vaidya and S.~K.~Vempati,
  Phys.\ Rev.\ D {\bf 62}, 093020 (2000)
  [hep-ph/0006138];
  A.~S.~Joshipura, R.~D.~Vaidya and S.~K.~Vempati,
  Nucl.\ Phys.\ B {\bf 639}, 290 (2002)
  [hep-ph/0203182];
  F.~de Campos, O.~J.~P.~Eboli, M.~B.~Magro, W.~Porod, D.~Restrepo, M.~Hirsch and J.~W.~F.~Valle,
  JHEP {\bf 0805}, 048 (2008)
  [arXiv:0712.2156 [hep-ph]];
  D.~Restrepo, M.~Taoso, J.~W.~F.~Valle and O.~Zapata,
  Phys.\ Rev.\ D {\bf 85}, 023523 (2012)
  [arXiv:1109.0512 [hep-ph]];
  F.~Bazzocchi, S.~Morisi, E.~Peinado, J.~W.~F.~Valle and A.~Vicente,
  JHEP {\bf 1301}, 033 (2013)
  [arXiv:1202.1529 [hep-ph]];
  E.~Peinado and A.~Vicente,
  Phys.\ Rev.\ D {\bf 86}, 093024 (2012)
  [arXiv:1207.6641 [hep-ph]];
  A.~Arhrib, Y.~Cheng and O.~C.~W.~Kong,
  arXiv:1210.8241 [hep-ph].

\bibitem{TestRp}
  W.~Porod, M.~Hirsch, J.~Romao and J.~W.~F.~Valle,
  Phys.\ Rev.\ D {\bf 63}, 115004 (2001)
  [hep-ph/0011248];
  M.~Hirsch, W.~Porod, J.~C.~Romao and J.~W.~F.~Valle,
  Phys.\ Rev.\ D {\bf 66}, 095006 (2002)
  [hep-ph/0207334];
  A.~Bartl, M.~Hirsch, T.~Kernreiter, W.~Porod and J.~W.~F.~Valle,
  JHEP {\bf 0311}, 005 (2003)
  [hep-ph/0306071];
  F.~de Campos, O.~J.~P.~Eboli, M.~B.~Magro, W.~Porod, D.~Restrepo, S.~P.~Das, M.~Hirsch and J.~W.~F.~Valle,
  Phys.\ Rev.\ D {\bf 86}, 075001 (2012)
  [arXiv:1206.3605 [hep-ph]];
  D.~Aristizabal Sierra, D.~Restrepo and S.~Spinner,
  arXiv:1212.3310 [hep-ph].


\bibitem{Carena-etal}
  M.~Carena, S.~Gori, N.~R.~Shah and C.~E.~M.~Wagner,
  JHEP {\bf 1203}, 014 (2012)
  [arXiv:1112.3336 [hep-ph]];
  M.~Carena, S.~Gori, N.~R.~Shah, C.~E.~M.~Wagner and L.~-T.~Wang,
  JHEP {\bf 1207}, 175 (2012)
  [arXiv:1205.5842 [hep-ph]].

\bibitem{Cao-etal}
  J.~Cao, Z.~Heng, T.~Liu and J.~M.~Yang,
  Phys.\ Lett.\ B {\bf 703}, 462 (2011)
  [arXiv:1103.0631 [hep-ph]];
  J.~-J.~Cao, Z.~-X.~Heng, J.~M.~Yang, Y.~-M.~Zhang and J.~-Y.~Zhu,
  JHEP {\bf 1203}, 086 (2012)
  [arXiv:1202.5821 [hep-ph]];
  J.~Cao, Z.~Heng, J.~M.~Yang and J.~Zhu,
  JHEP {\bf 1210}, 079 (2012)
  [arXiv:1207.3698 [hep-ph]].

\bibitem{Arbey-etal}
  A.~Arbey, M.~Battaglia, A.~Djouadi and F.~Mahmoudi,
  JHEP {\bf 1209}, 107 (2012)
  [arXiv:1207.1348 [hep-ph]].

\bibitem{NMSSM}
  U.~Ellwanger,
  JHEP {\bf 1203}, 044 (2012)
  [arXiv:1112.3548 [hep-ph]];
  K.~Schmidt-Hoberg and F.~Staub,
  JHEP {\bf 1210}, 195 (2012)
  [arXiv:1208.1683 [hep-ph]].

\bibitem{Djouadi}
  A.~Djouadi,
  Phys.\ Rept.\  {\bf 459}, 1 (2008)
  [hep-ph/0503173].

\bibitem{Kita}
  T.~Kitahara,
  JHEP {\bf 1211}, 021 (2012)
  [arXiv:1208.4792 [hep-ph]];
  T.~Kitahara, T.~Yoshinaga and ,
  arXiv:1303.0461 [hep-ph].

\bibitem{pdg}
  J. Beringer {\it et al.} (Particle Data Group), Phys.\ Rev.\ D {\bf 86}, 010001 (2012).

\bibitem{theta13}
  Y.~Abe {\it et al.}  [DOUBLE-CHOOZ Collaboration],
  Phys.\ Rev.\ Lett.\  {\bf 108}, 131801 (2012)
  [arXiv:1112.6353 [hep-ex]];
  F.~P.~An {\it et al.}  [DAYA-BAY Collaboration],
  Phys.\ Rev.\ Lett.\  {\bf 108}, 171803 (2012)
  [arXiv:1203.1669 [hep-ex]];
  J.~K.~Ahn {\it et al.}  [RENO Collaboration],
  Phys.\ Rev.\ Lett.\  {\bf 108}, 191802 (2012)
  [arXiv:1204.0626 [hep-ex]].

\bibitem{glob-fit}
  D.~V.~Forero, M.~Tortola and J.~W.~F.~Valle,
  arXiv:1205.4018 [hep-ph].

\bibitem{tribi}
  P.~F.~Harrison, D.~H.~Perkins and W.~G.~Scott,
  Phys.\ Lett.\ B {\bf 530}, 167 (2002)
  [hep-ph/0202074].

\bibitem{Hundi}
  R.~S.~Hundi,
  Phys.\ Rev.\ D {\bf 83}, 115019 (2011)
  [arXiv:1101.2810 [hep-ph]].

\bibitem{Hdecay}
  A.~Djouadi, J.~Kalinowski and M.~Spira,
  Comput.\ Phys.\ Commun.\  {\bf 108}, 56 (1998)
  [hep-ph/9704448].

\bibitem{mug-2}
For a review on the muon $(g-2)$, see,
  Z.~Zhang,
  arXiv:0801.4905 [hep-ph];
  F.~Jegerlehner and A.~Nyffeler,
  Phys.\ Rept.\  {\bf 477}, 1 (2009)
  [arXiv:0902.3360 [hep-ph]].

\bibitem{MSSMg-2}
  T.~Moroi,
  Phys.\ Rev.\ D {\bf 53}, 6565 (1996)
  [Erratum-ibid.\ D {\bf 56}, 4424 (1997)]
  [hep-ph/9512396];
  S.~P.~Martin and J.~D.~Wells,
  Phys.\ Rev.\ D {\bf 64}, 035003 (2001)
  [hep-ph/0103067].

\bibitem{Polonyi}
J. Polonyi, Hungary Central Research Institute Report
No. KFKI-77-93, 1977 (unpublished).

\bibitem{Kim-Nilles}
  J.~E.~Kim and H.~P.~Nilles,
  Phys.\ Lett.\ B {\bf 138}, 150 (1984).

\end{thebibliography}
\end{document}